\documentclass[fleqn,usenatbib]{mnras}

\usepackage{newtxtext,newtxmath}

\usepackage[T1]{fontenc}

\DeclareRobustCommand{\VAN}[3]{#2}
\let\VANthebibliography\thebibliography
\def\thebibliography{\DeclareRobustCommand{\VAN}[3]{##3}\VANthebibliography}

\usepackage{graphicx}	
\usepackage{amsmath}	
\usepackage{xcolor}

\title[Hybrid Galaxy Morphology]{Galaxy Morphology in CANDELS: Addressing Evolutionary Changes Across $0.2 \leq z \leq 2.4$ with Hybrid Classification Approach}

\author[Kolesnikov et al. 2024]{
I. Kolesnikov,$^{1}$\thanks{E-mail: igor.s.kolesnikov@gmail.com}
V. M. Sampaio,$^{2}$
R. R. de Carvalho,$^{2,4}$
C. Conselice,$^{1}$
\\
$^{1}$ Jodrell Bank Centre for Astrophysics, University of Manchester, Oxford Road, Manchester M13 9PL, UK\\
$^{2}$ NAT - Universidade Cidade de São Paulo, 01506-000, SP, Brazil\\
$^{3}$ Universidade de São Paulo, IAG, Rua o Matão 1226, Cidade Universitária, São Paulo 05508-900, Brazil\\
}

\date{Accepted XXX. Received YYY; in original form ZZZ}

\pubyear{2024}

\begin{document}
\label{firstpage}
\pagerange{\pageref{firstpage}--\pageref{lastpage}}
\maketitle

\begin{abstract}
Morphological classification of galaxies becomes increasingly challenging with redshift. We apply a hybrid supervised-unsupervised method to classify $\sim 14,000$ galaxies in the CANDELS fields at $0.2 \leq z \leq 2.4$ into spheroid, disk, and irregular systems. Unlike previous works, our method is applied to redshift bins of width 0.2. Comparison between models applied to a wide redshift range versus bin-specific models reveals significant differences in galaxy morphology beyond $z > 1$ and a consistent $\sim 25\%$ disagreement. This suggests that using a single model across wide redshift ranges may introduce biases due to the large time intervals involved compared to galaxy evolution timescales. Using the FERENGI code to assess the impact of cosmological effects, we find that flux dimming and smaller angular scales may lead to the misclassification of up to 18\% of disk galaxies as spheroids or irregulars. Contrary to previous studies, we find an almost constant fraction of disks ($\sim 60\%$) and spheroids ($\sim 30\%$) across redshifts. We attribute discrepancies with earlier works, which suggest a decreasing fraction of disks beyond $z = 1$, to the biases introduced by visual classification. Our claim is further strengthened by the striking agreement to the results reported by Lee et al. (2024) using an objective, unsupervised method applied to James Webb Space Telescope data. Exploring mass dependence, we observe a $\sim 40\%$ increase in the fraction of massive ($ \rm M_{\rm stellar} \geq 10^{10.5} M_{\odot}$) spheroids with decreasing redshift, well balanced with a decrease of $\sim 20\%$ in the fraction of $\rm M_{\rm stellar} \geq 10^{10.5}$ disks, suggesting that merging massive disk galaxies may form spheroidal systems.

\end{abstract}



\section{Introduction}

Morphology has been a fundamental probe of galaxy formation and evolution. Morphological features reveal dynamical mechanisms shaping galaxies \citep{1969ApJ...155..393P, 2015ApJ...812...29T}. For example, high angular momentum in a protogalactic cloud leads to the formation of a rotating disk structure, resulting in spiral galaxies with well-defined disks and spiral arms. Conversely, low angular momentum causes a more isotropic collapse, forming spheroidal structures typical of elliptical galaxies. Irregular distributions of angular momentum can give rise to morphologies in between spheroidal and disk galaxies, often referred as irregular systems.

Morphology is also connected to variations in star formation efficiency. In disk galaxies, differential rotation promotes star formation, particularly in spiral arms where density waves compress the gas \citep{2019igfe.book.....C}. In spheroidal galaxies, rapid gas collapse can lead to intense, but short-lived star formation, resulting in older stellar populations in present-day observations and negligible ongoing star formation \citep{Strateva1, 2012MNRAS.424..232W, 2020MNRAS.491.5406T}.

Furthermore, the morphology–density and morphology–radius relations indicate a higher proportion of early-type galaxies (ellipticals and S0 galaxies) in denser regions of galaxy clusters and closer to cluster centers compared to low-density fields \citep{Dressler, 1997ApJ...490..577D}, indicating that environment can also leave significant imprints in galaxy morphology. In the local universe, data from the Sloan Digital Sky Survey \citep[SDSS,][]{2000AJ....120.1579Y} have shown that elliptical galaxies are predominantly found at high stellar masses and occupy specifically the red sequence region in the star formation main sequence diagram, whereas spiral and S0 galaxies are mainly found in the blue cloud green valley, respectively \citep{2014MNRAS.440..889S, 2023MNRAS.524.5327S}.

\begin{figure*}
	\includegraphics[width=505px]{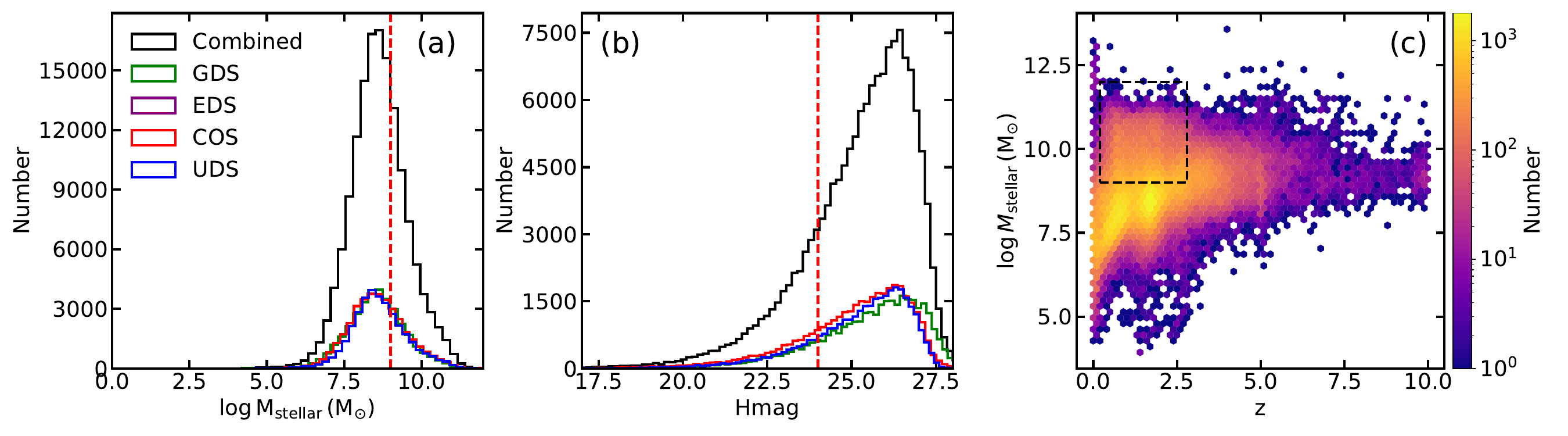}
    \caption{Panel (a) displays the median stellar mass distribution across each field, whereas Panel (b) presents the distribution of apparent magnitudes in the H band. Red lines demonstrate cuts we applied to select data for our sample. Panel (c) shows the distribution of stellar masses of CANDELS fields with selected region in black rectangle.}
    \label{fig:mass_mag_candels}
\end{figure*}

At higher redshifts, the Hubble Space Telescope (HST) is a crucial tool for characterizing galaxy morphology due to its high resolution and broad wavelength coverage from optical to near-infrared. Studies using HST data have consistently reported that galaxies at $z \sim 2$ often exhibit disturbed or irregular features, suggesting a predominance of irregular galaxies at early stages of the universe \citep{2014ARA&A..52..291C}. However, reliably characterizing galaxy morphology at high redshifts is challenging due to cosmological effects and observational limitations. For instance, decreasing angular size and flux dimming with redshift. Different methods yield significantly different results. For instance, \citet{2012MNRAS.427.1666B} characterize the Sérsic index of massive galaxies ($\log(\rm M_{stellar}/{\rm M}_{\odot}) \geq 11$) and find a significant increase in disk-like galaxies from 15\% at $z = 0$ to 80\% at $z = 2$, while the fraction of spheroidal galaxies decreases from 85\% to 20\% over the same range. In contrast, \citet{mortlock2013redshift} visually classify galaxies with $\log(\rm M_{stellar}/{\rm M}_{\odot}) \geq 10$ at $z > 2$ from the UltraDeep Survey (UDS) and find a higher fraction of spheroidal galaxies (40\%) compared to disk-like systems (10\%). These discrepancies highlight the impact of different classification methods on morphological analyses at high redshift.

Parametric approaches, such as light profile fitting, depend on well-behaved light distributions, which are often not the case at high redshifts. For instance, fitting a Sérsic law to an irregular galaxy light profile is senseless, as it has subtleties not taken into account in the adopted law.  
On the other hand, non-parametric approaches, like the Concentration + Asymmetry + Smoothness (CAS) system, does not rely on any assumption about the galaxy light profile, thus being able to characterize even irregular galaxies \citep{barchi2020machine}. However it has been shown that this system has a limited performance in distinguishing between disk and bulge dominated galaxies. As an alternative, \citet{kolesnikov2023unveiling} shows that a system combining Entropy + Gini Index + Gradient Patter Asymmetry (EGG) is more reliable in separating these two morphological types. Nevertheless,  whichever approach is used, pre-processing has always an effect either on the fitting of the light profile or the measurement of the metrics and, as we progress to higher redshifts, less pixels are available for analysis.

Alongside parametric and non-parametric methods, machine learning (ML) and deep learning (DL) algorithms are gaining more attention each year. The benefits of these tools, such as scalability, speed, and accuracy, provide substantial improvements in classification and analysis. Multiple papers use ML/DL algorithms to achieve automated classifications \citep{ferrari2015morfometryka, dominguez2018improving, barchi2020machine, khalifa2017deep, primack2018deep, khan2019deep, tohill2021quantifying, walmsley2022galaxy, cheng2023lessons, walmsley2023galaxy}. More recently,  works \citep{popp2024transfer, wei2023galaxy} apply transfer learning and backbone augmentation with attention and deformable convolution. Such methods offer numerous benefits \citep{zhao2023comparison, Bello_2019_ICCV, Dai_2017_ICCV}. However, sophisticated methodologies will only be effective if the ground truth catalogs are well defined, since, being a supervised approach, the model will inherit all imperfections of the training set.

Currently, one way to define ground truth datasets is through citizen science projects such as Galaxy Zoo \citep{lintott2011galaxy} or Zooniverse \citep{simpson2014zooniverse}. In addition, small teams of astronomers can classify smaller amounts of data with specific goals in mind \citep{kartaltepe2015candels, cassata2005evolution, mortlock2013redshift, talia2014listening, delgado2010hubble}. In both scenarios, the classification is done by visual inspection. While the human eye possesses a great capacity for integration and feature extraction, it is also susceptible to bias and subjectivity, where small, faint, or distant galaxies could be classified differently by different individuals. In this paper, we argue that visual bias has a serious impact on morphological analysis due to its propagation to ML models, especially at high redshifts.

Combining these two critical points, bias and time consumption, we note a gap that needs to be filled by a method that will be fast, general, objective, reproducible, and applicable to different datasets. In this work, we introduce a novel method to classify galaxy morphology at $0.2 \leq z \leq 2.4$ using a hybrid supervised-unsupervised approach. We apply this method to galaxies from the Cosmic Assembly Near-IR Deep Extragalactic Legacy Survey (CANDELS) observed in the F814W filter by the HST Wide Field Camera (WFC). We implement updates in our method in comparison to \citet[][K24 hereon]{kolesnikov2023unveiling}, in order to provide reliable classification for high redshift galaxies. We also examine the biases introduced by using a single model across a wide redshift range and investigate the fraction of disk and spheroidal galaxies as a function of redshift, comparing our findings with previous studies.

This paper is organized as it follows: in Section 2, we present the sample selection and the catalogs used to retrieve galaxy properties; in Section 3, we provide a brief description of the hybrid unsupervised-supervised method and present some modifications with respect to the methodology employed in K24; in Section 4, we present a detailed investigation on the performance of CNN classification due to redshift-related effects; in Section 5, we present our results and discuss observed trends in a cosmological context; and Section 6, summarizes the main results of this work. Through this paper we adopt, we adopt a flat $\Lambda$CDM cosmology with $\rm [\Omega_{M}, \Omega_{\Lambda}, H_{0}] = [0.31, 0.69, 68 \, {\rm km \, s^{-1} \, Mpc^{-1}}]$ to be consistent with \textit{Planck 18 cosmology} \citep{2020A&A...641A...6P} and report the magnitudes in the AB system.

\section {Sample Selection and Data Used}\label{sec:data}

In this paper, we utilize a combination of fields from the Cosmic Assembly Near-infrared Deep Extragalactic Legacy Survey (CANDELS) \citep{faber2011cosmic, koekemoer2011candels, grogin2011candels} as our primary data sample. CANDELS is a 902-orbit legacy program designed to study galaxy formation and evolution over a wide redshift range using the near-infrared WFC3 camera on the Hubble Space Telescope, which obtains deep imaging of faint and distant objects. To date, CANDELS has imaged over 250,000 distant galaxies within five strategic regions: The Great Observatories Origins Deep Survey South and North (GOODS-S and GOODS-N), Ultra-Deep Survey (UDS), Extended Groth Strip (EGS), and Cosmic Evolution Survey (COSMOS), covering a combined area of approximately $0.22$ deg\(^2\) \citep[e.g.][]{barro2019candels}.

We restrict our sample to the COSMOS, EGS, UDS, and GOODS-S fields, due to the uniformity in photometric catalog and methodology to estimate stellar masses \citep{santini2015stellar, stefanon2017candels, nayyeri2017candels}. In a few words, stellar mass estimates are derived by SED fitting the observed multiwavelength photometry (ranging from infrared to ultraviolet) with stellar population synthesis templates \citep{1997A&A...326..950F, 2003MNRAS.344.1000B, 2005MNRAS.362..799M, 2007ASPC..374..303B}. To avoid biases due to the choice of minimization method and/or adopted fitting code, the effort of 10 different teams is combined, each one adopting their preferred fitting code, assumptions, priors, and parameter grid. Therefore, the final results using the same underlying stellar isochrones reduces the systematic uncertainties associated with the fitting code and stellar population choice. Although the original CANDELS photometric catalog classifies galaxies using the F160W filter ($\lambda_{\rm eff}^{\rm F160W} = 14,445$\AA), we select only galaxies that were observed using the F814W filter ($\lambda_{\rm eff}^{\rm F814W} = 8050$\AA) from the Advanced Camera for Surveys (ACS). This follows from a better resolution available for F814W observations, namely a pixel scale of 0.05"/px, in comparison to 0.1"/px for the WFC3.

In Figure \ref{fig:mass_mag_candels}, we present stellar mass and apparent magnitude distributions in panels (a) and (b), respectively, for each field (different color) and when combined (in dark). Notably, the distributions are similar for all the four fields, thus we apply the same thresholds irrespective of field. First, we select only galaxies with stellar masses greater than $10^9 {\rm M}_\odot$ to avoid including dwarf galaxies in our sample, since these kind of objects have a higher uncertainty associated to SED fitting \citep{santini2015stellar}. Second, we limit our sample to galaxies brighter than $\rm H = 24$. We highlight that, despite utilizing ACS F814W images, we apply magnitude thresholds based on the F160W filter, which was originally used to detect and characterize galaxies in the four fields \citep{barro2019candels}. The value $\rm H = 24$ is a conservative threshold that follows from works suggesting that both non-parametric morphological indices and automated algorithms struggle to reliably characterize galaxy morphology for fainter systems \citep{grogin2011candels,kartaltepe2015candels}. In panels (a) and (b), the red vertical line shows the adopted thresholds. In panel (c), we show the number of objects as a function of stellar mass and redshift for the four fields combined. Notably, there is a significant decrease in the number of objects more massive than $10^{10}{\rm M}_{\odot}$ above $z \sim 2.4$. Therefore, we limit our analysis to galaxies in $0.2 \leq z \leq 2.4$. Our combined selection criteria are illustrated by the dashed black rectangle, which comprises 16,718 galaxies available in F814w ACS/WFC.

\subsection{Data Preparation Strategy}

\begin{figure*}
	\includegraphics[width=505px]{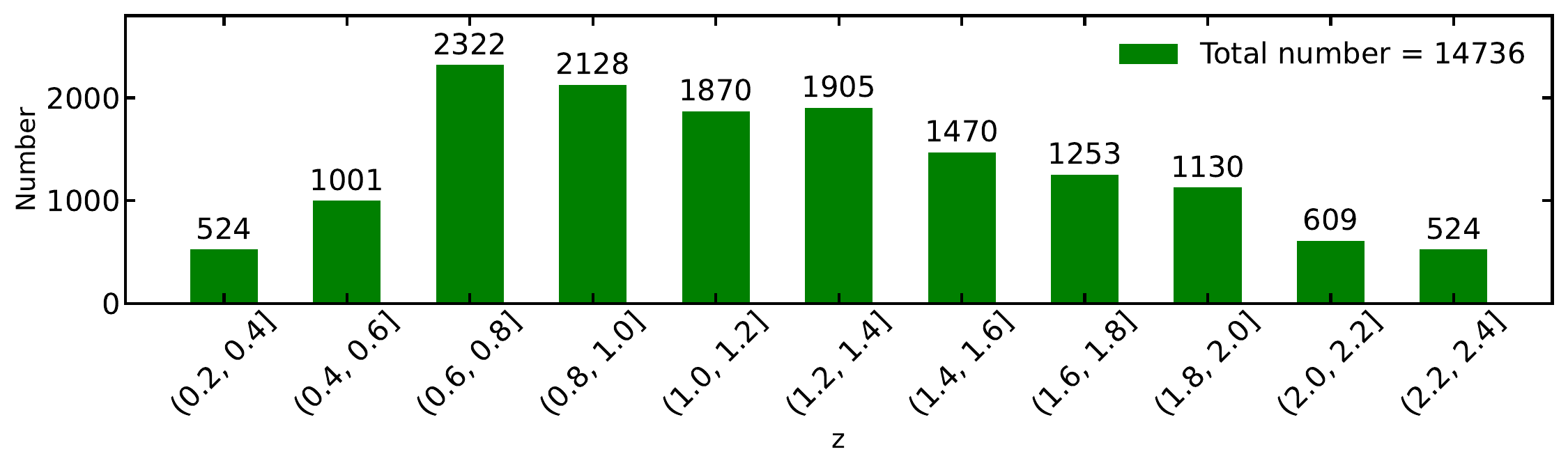}
    \caption{Final galaxy selection for hybrid classification after applying the cuts of $\rm Hmag = 24$ and stellar mass greater than $10^9 {\rm M}_\odot$, followed by cleaning, segmentation of cutouts, and metric measurement. The data is divided into redshift bins, with each bin containing galaxies within a specified redshift range, treated as an independent dataset.}
    \label{fig:base_qtd_combined}
\end{figure*}

As outlined in K24, data preparation is a fundamental step, especially when estimating the non-parametric indices. Using the combined drizzled F814W fields provided by the CANDELS survey\footnote{All CANDELS fields can be accessed at \url{https://archive.stsci.edu/hlsp/candels}}, we create cutouts centered in each galaxy of the sample and with size empirically defined as $\rm K \times FLUX\_RADIUS$ (in pixels), where $\rm FLUX\_RADIUS$ is the radius containing 50\% of the galaxy total light in the F814W band. We use $\rm K=5$ for cutouts used for metric extraction and $K=3$ for classification using DL. The difference is justified by the need of background subtraction during metric extraction, which require a wide ``empty'' area around the galaxy. On the other hand, for the DL classification, only the galaxy image is needed.  
We remove 1,982 galaxies from our sample, as they are located close to the CANDELS follow-up edges or suffer from the presence of nearby bright objects, which may input bias in the non-parametric indexes estimation. Even though we remove 11\% of the total number of galaxies, we ensured that this does not alter the original stellar mass and Hmag distribution. After this exclusion, our sample consists of 14,736 galaxies. We further separate galaxies into redshift bins varying from 0.2 to 2.4 in steps of 0.2, where the step is selected to guarantee at least 500 galaxies in each bin. We show the number of galaxies in each redshift bin in Figure~\ref{fig:base_qtd_combined}. This separation in redshift bins is particularly relevant to our method, which is described in details in Section~\ref{method_update}.

Using the original cutout for each galaxy, we perform background subtraction, cleaning and segmentation mask. The last is particularly relevant as it defines the pixels belonging to the galaxy, thus the ones that are used in non-parametric indices measurements. Since our sample covers a wide redshift range, we define our segmented mask as the best fitting ellipse with semi major axis equal to the Petrosian radius, which is irrespective of redshift.

\subsection{The Morphological System $\texttt{M}_\texttt{2O}$ + \texttt{EGG} in a Nutshell}

In this work we use the the EGG system as a first approach to characterize galaxies morphology. This choice follows from a better performance of this system in comparison to the commonly CAS system adopted in the literature (e.g. K24). In addition, we include the second moment of light distribution $\texttt{M}_\texttt{2O}$, defining the \texttt{MEGG} system. The inclusion of $\texttt{M}_\texttt{2O}$ follows from promising results with SDSS and HST morphological separations in \citet{lotz2004new} and \citet{cheng2021galaxy}. Moreover, the additional metric results in a more refined set of non-parametric indexes that may help to define better and unbiased training samples (See Section \ref{method_update}). Below we briefly describe the metrics composing the \texttt{MEGG} system, emphasizing that we estimate the metrics for each galaxy in the best fitting ellipse with semi major axis equal to the Petrosian radius, in order to guarantee robustness irrespective of redshift.

\begin{itemize}
    \item \textbf{Entropy (\texttt{E}):} The entropy of information characterizes the distribution of pixel values in an image. Originating from Shannon entropy, it captures the randomness or unpredictability inherent in the images' information content \citep{bishop2007pattern, ferrari2015morfometryka}. Following this definition, spheroidal galaxies have lower entropy in comparison to disk systems.
    
    \item \textbf{Gini Coefficient (\texttt{G}):} Originally used in economics to represent wealth distribution, the G coefficient has been adapted for galaxy morphology to measure the relative flux distribution across pixels corresponding to a galaxy. While it correlates with concentration, it does not necessarily presume the brightest pixels to be centrally positioned in the galaxy image \citep{abraham2003}. Therefore, a high Gini index indicates a large concentration of light in a small number of pixels, which is characteristic of spheroidal galaxies. Conversely, a low Gini suggests a more even distribution of light across the galaxy, which is mainly seen in disk systems.
    
    \item \textbf{Second Gradient Moment ($\texttt{G}_\texttt{2}$}): Gradient Pattern Analysis (GPA) is a method crafted to gauge gradient bilateral asymmetries in a numerical grid. Within galaxy morphometry, the second gradient moment, represented as $\texttt{G}_\texttt{2}$, stands out. It proves especially valuable, showcasing its potential to differentiate spheroids from disk systems more effectively than traditional morphometric metrics. For instance, $\texttt{G}_\texttt{2}$ achieves about 90\% accuracy in separating galaxies using data from the SDSS-DR7 catalog \citep{rosa2018gradient}. The computation of $\texttt{G}_\texttt{2}$ starts by calculating the gradient field in the X and Y directions of the image. With these two components at each location, we attribute a vector (characterized by module and phase) for each pixel. Then, we compare the magnitude and direction of equidistant vectors from the galaxy center. Pairs that are symmetric (same magnitude but opposite direction) are excluded, resulting in a field with only asymmetric vectors. The next step involves calculating the number of and the sum of the asymmetric vectors (See K24 for more details on how $\texttt{G}_\texttt{2}$ is estimated.
    
    \item \textbf{The second moment of light  ($\texttt{M}_\texttt{2O}$}):  The $\texttt{M}_\texttt{2O}$ parameter is calculated from the second-order moment of the brightest regions of a galaxy. The total second-order moment, $\rm M_{\text{tot}}$, of a galaxy's light distribution is defined as:
    \begin{equation}
    \rm M_{\text{tot}} = \sum_{i} M_i = \sum_{i} f_i \left[ (x_i - x_c)^2 + (y_i - y_c)^2 \right],    
    \end{equation}
    where $f_i$ is the flux in the $i$-th pixel, $(x_i, y_i)$ are the coordinates of the $i$-th pixel, and $(x_c, y_c)$ are the coordinates of the galaxy's center.
    
    The $\texttt{M}_\texttt{2O}$ parameter is then defined as the normalized second-order moment of the brightest 20\% of the galaxy's flux,
    \begin{equation}        
    \rm M_{20} = \log \left( \frac{\sum_{i} M_i}{M_{\text{tot}}} \right), \text{while } \sum f_{i} \leq 0.2 F_{\rm tot}
    \end{equation}
    where $\rm F_{\text{tot}}$ is the total flux of the galaxy, and $\rm f_{i}$ is the flux in a given pixel. To calculate $\texttt{M}_\texttt{2O}$, we follow: 1) rank all pixels by their flux values in descending order; 2) calculate the cumulative flux $\rm F_{\text{cum}}$ until it reaches 20\% of the total flux $\rm F_{\text{tot}}$; 3) sum the second-order moment $\rm M_i$ for all pixels contributing to this cumulative flux; and 4) normalize this sum by the total second-order moment $\rm M_{\text{tot}}$. With this definition, lower values of $\texttt{M}_\texttt{2O}$ indicate that the brightest regions are more centrally concentrated -- related to spheroidal galaxies, while higher values suggest a more extended flux distribution -- suggesting disk galaxies.
\end{itemize}

\section{An Updated Version of the Hybrid Approach}\label{method_update}

In this section, we present the method and processing pipeline, along with its updates and refinements in comparison to K24. It is important to note that extracting morphological information for the galaxies in this high-z sample is significantly more challenging due to the degradation effect (which is explored in Section \ref{redshift}). For this reason, we revise some steps in our method. 

\begin{figure*}
	\includegraphics[width=505px]{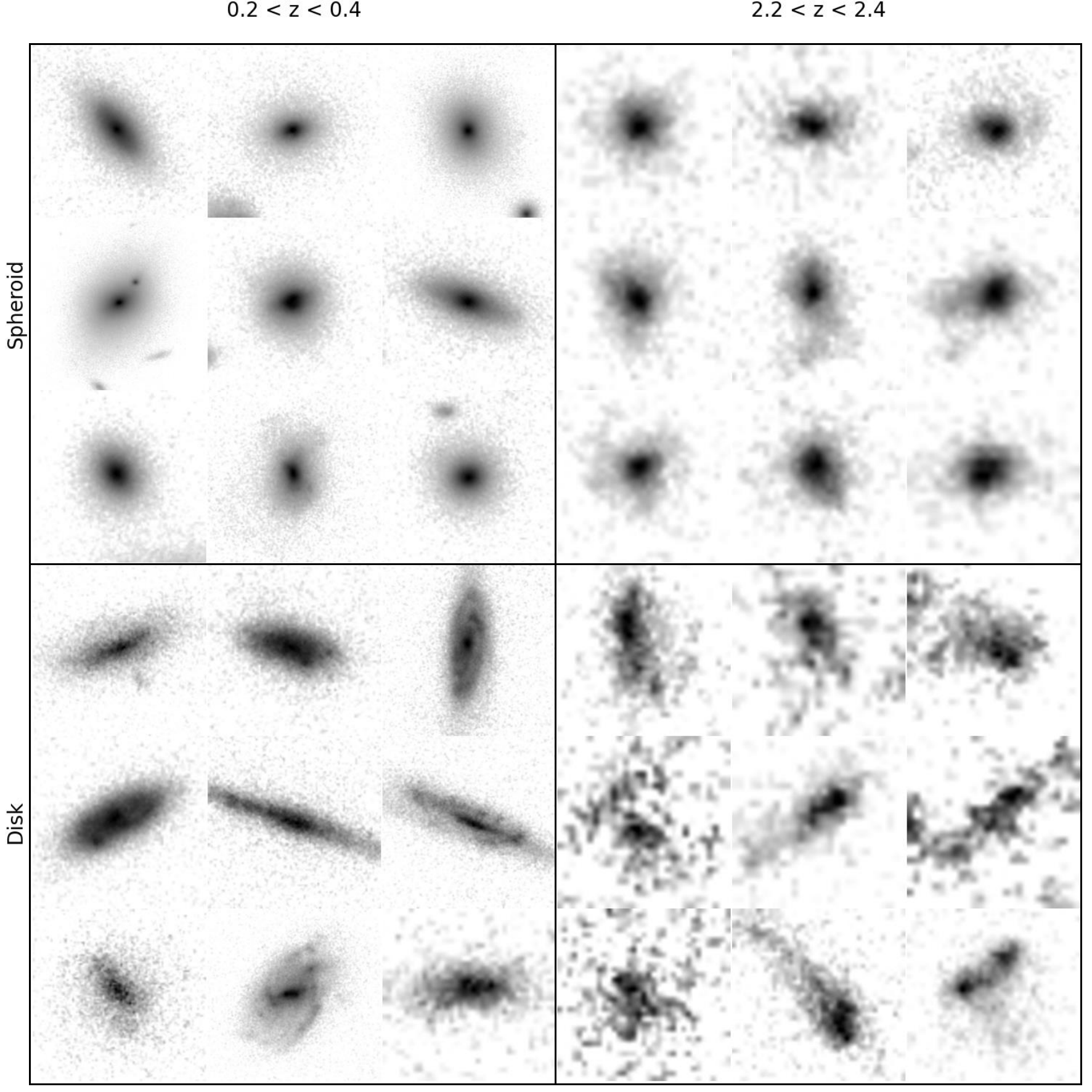}
    \caption{Examples of randomly selected galaxies from out dataset basing on our model classification. This figure is split into 4 segments. Top Left 3x3 segment is showing 9 galaxies classified as spheroidal at 0.2 < z < 0.4 (first bin); Top Right 3x3 segment is showing 9 galaxies classified as spheroidal at 2.2 < z < 2.4 (last bin); Bottom Left 3x3 segment is showing 9 galaxies classified as disk at 0.2 < z < 0.4 (first bin); Bottom Right 3x3 segment is showing 9 galaxies classified as disk at 2.2 < z < 2.4 (last bin). We note how much different these higher redshift galaxies look like, especially in case of disks and how much more complex get the task of the visual classification in higher redshifts.}
    \label{fig:main_mosaic}
\end{figure*}

\subsection{SOMbrero and Unsupervised Labeling}

We use the non-parametric indices vector (${v}$ = [$\texttt{M}_\texttt{2O}$, \texttt{E}, \texttt{Gini}, $\texttt{G}_\texttt{2}$]) as input for a Self-Organizing Maps (SOM) clustering algorithm, implemented through the SOMbrero R-package \citep{villa2017stochastic}. We choose SOMbrero primarily due to its ability to process numeric datasets, given that we supply the metrics as numerical quantifiers of galaxies' physical properties. Moreover, the SOMbrero package offers a robust functional foundation for working with data flow, distinguishing it from other implementations and clustering methods. As pointed out in K24, without meticulous cluster selection, SOMbrero demonstrates classification performance comparable to other clustering algorithms.

Initially, all galaxies in our dataset are unlabeled, and the description of morphology comes from the metrics. SOMbrero then defines a set o $\rm N_{\rm clusters}$, each initialized with a random value for each metric. For each galaxy, SOMbrero calculates the Euclidean distance between the metrics vector $\rm \vec{v}$ and the i-th cluster weight vector $\rm\vec{w_{i}}$. The cluster with the smallest distance is identified as the Best Matching Unit (BMU), and the weights are adjusted to become closer to the galaxy's metrics, following the equation
\begin{equation}
\rm \mathbf{w}(t+1) = \mathbf{w}(t) + \theta(t, r) \cdot \alpha(t) \cdot (\mathbf{v} - \mathbf{w}(t)),    
\end{equation}
    
where \(\rm \mathbf{w_{i}}(t) \) is the weight vector of the cluster at step \(\rm t \), \( \rm\mathbf{w}(t+1) \) is the updated weight vector, \(\rm \mathbf{v} \) is the metrics vector of the current galaxy, \(\rm \alpha(t) \) is the learning rate -- i.e. a quantity that tells us how much the weights change in response to each galaxy, and typically decreases over time, \(\rm \theta(t, r) \) is the neighborhood function (often a Gaussian) -- that defines how the update of weights decreases for nodes farther from the BMU, and decreases with BMU distance and time, and \( r \) is the radius of the neighborhood around the BMU affected by the update.

In early stages, weights (higher \(\rm \alpha(t) \) and larger neighborhood radius \( r \)) suffer larger changes, allowing the SOM to organize itself according to the data's structure. As the training progresses, \(\rm \alpha(t) \) decreases and \(\rm r \) may also decrease, leading to finer adjustments and stabilization of the map. In our pipeline, we use the Letremy implementation of distance and radius calculations. The Letremy distance is a type of measure designed to handle data by measuring the dissimilarity between two variables. It takes into account the co-occurrence frequencies of categories across the dataset.

More details can be found in the original publications \citep{cottrell2004som, cottrell2005use}.

The final result of the SOM algorithm is a grid of clusters, each containing galaxies with similar metric distributions. The clusters that demonstrate the most defined metric distribution generally contain the most distinct objects (See Figure 12 of K24) . We term these clusters as ``prominent clusters'' and base our label assignment on them.

\subsection{Prominent Cluster Selection}

In comparison to K24, the prominent cluster selection process has undergone slight changes to enhance robustness and flexibility. Instead of using the median with a shift to select prominent clusters based on each metric, we implemented the IsoData (Iterative Self-Organizing Data Analysis Technique Algorithm) clustering algorithm \citep{velasco1979thresholding, lloyd1982least, jensen1996introductory} that finds the threshold for each metric. The IsoData algorithm is particularly useful for scenarios where the number of clusters and their initial centroids are not known a priori, allowing the algorithm to dynamically adjust these parameters during its execution. In our case, we use the IsoData algorithm to analyze and partition the data based on a specified metric within a DataFrame. This partition signifies the distinction between the two morphological classes, determining at which value a given metric indicates whether a galaxy is classified as a disk or spheroidal.

We run SOMbrero five times with variable grid sizes (calculated as $\pm 2$ from the ideal grid size ($\rm L_{grid}$) calculated by 
\begin{equation}
    \rm L_{grid} = \sqrt{N_{\rm objects} \times 0.1},
\end{equation}
where $\rm N_{objects}$ is the number of objects inputted to the SOM algorithm. Prominent clusters are then selected from each run, and only galaxies consistently assigned to the same supercluster in all runs are preserved and assigned the respective label -- spheroidal or disk. We show in Figure~\ref{fig:main_mosaic} examples of galaxies from prominent clusters. We select 4 sets of 9 galaxies each, distributed as follows: top left and right -- 18 galaxies (9 each) classified as spheroidal in the redshift ranges $0.2 \leq z < 0.4$ and $2.2 \leq z < 2.4$, respectively; bottom left and right, same as top panels, but showing systems classified as disks.

\subsection{Unsupervised Optimization}\label{sec:unsupervised_optimization}

Two out of the four metrics have tunable parameters, which can affect their performance in separating disk and spheroidal galaxies. While $\texttt{M}_\texttt{2O}$ and \texttt{G} do not contain any tunable parameters, $\texttt{G}_\texttt{2}$ depends on module tolerance ($\rm m_{tol}$) and phase tolerance ($\rm p_{tol}$), and \texttt{E} depends on the number of bins ($\rm n_{bins}$) (See Section 3.2 of K24). In order to avoid making assumptions about the underlying distributions, we adopt an unsupervised method to define $\rm m_{tol}$, $\rm p_{tol}$ and $\rm n_{bins}$ that best discriminate galaxies according to their morphological features. 

\begin{figure}
	\includegraphics[width=\columnwidth]{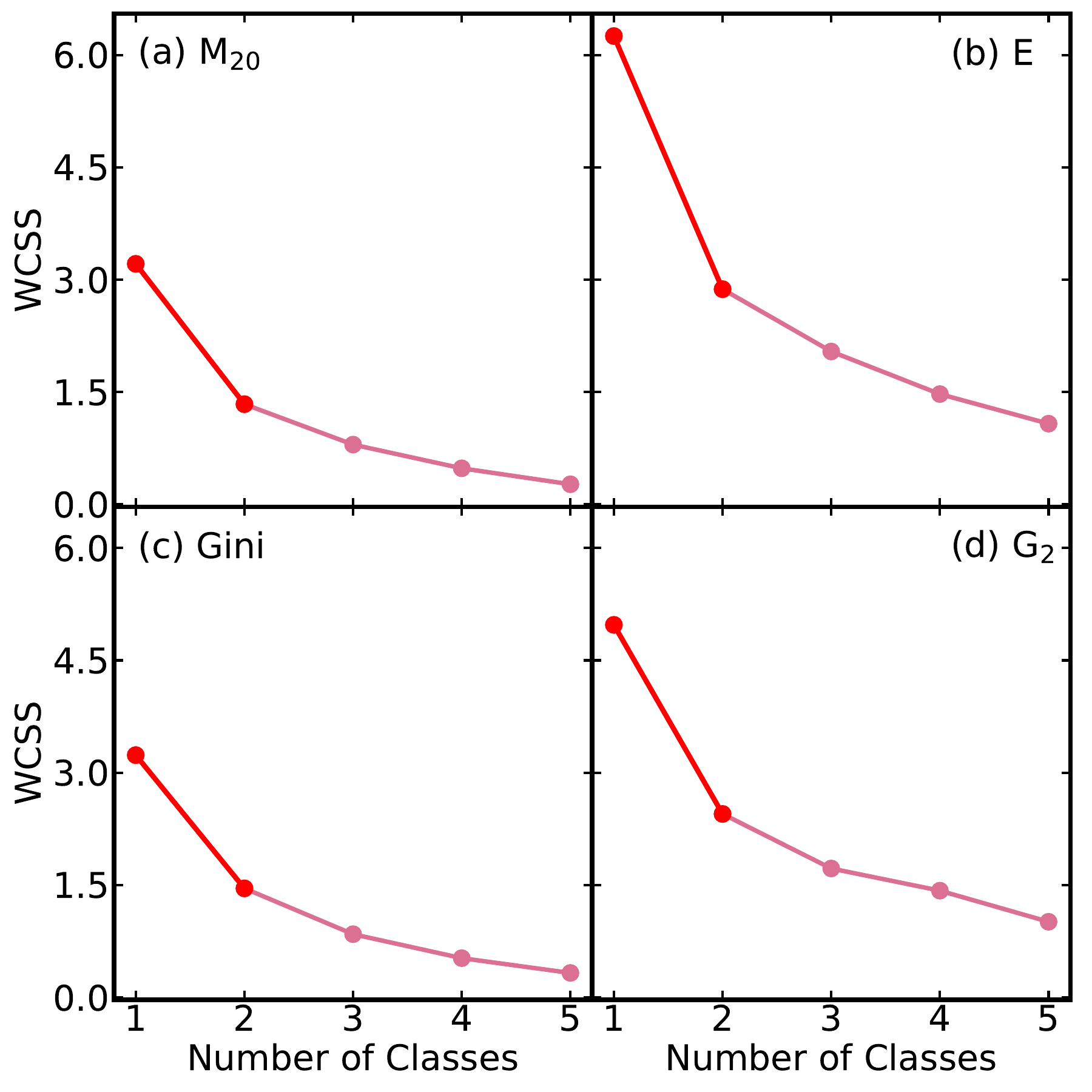}
    \caption{Comparison between metric capacity to split galaxies into multiple classes, at the same time indicating the optimal number of supercluster for each metric (signalized by red segment). It is possible to note that slope of red segment in each panel a is much steeper, indicating the higher change in WCSS, which in its turn points to a better super cluster split.}
    \label{fig:wcss}
\end{figure}

For this optimization, we use the ''knee value'' obtained from SOMbrero clustering. The method in question utilizes the clustering algorithm, which naturally aims at discriminating galaxies with distinct morphological features. To optimize each parameter, $\rm m_{tol}$, $\rm p_{tol}$, and $\rm n_{bins}$,  we test a range of possible values for these parameters. We then run the optimization for each value from this list and select the one corresponding to the highest ''knee value'' -- defined as the steepest fall of the within-cluster sum of squares (WCSS) given by:
\begin{equation}
\rm \text{WCSS} = \sum_{i=1}^{N} \sum_{j=1}^{d} \left(x_{ij} - \mu_{c(i)j}\right)^2    
\end{equation}

Where:
\begin{itemize}
    \item  \(\rm N \): Total number of data points (individual galaxies).
    \item  \(\rm d \): Number of dimensions (features) of the data (metrics).
    \item  \(\rm x_{ij} \): Value of the \(\rm j \)-th feature of the \(\rm i \)-th data point.
    \item  \(\rm \mu_{c(i)j} \): Value of the \(\rm j \)-th feature of the center (mean) of the cluster \(\rm c(i) \) to which the \(\rm i \)-th data point belongs.
    \item  \(\rm c(i) \): Cluster assignment of the \(\rm i \)-th data point.
\end{itemize}

\begin{figure}
	\includegraphics[width=\columnwidth]{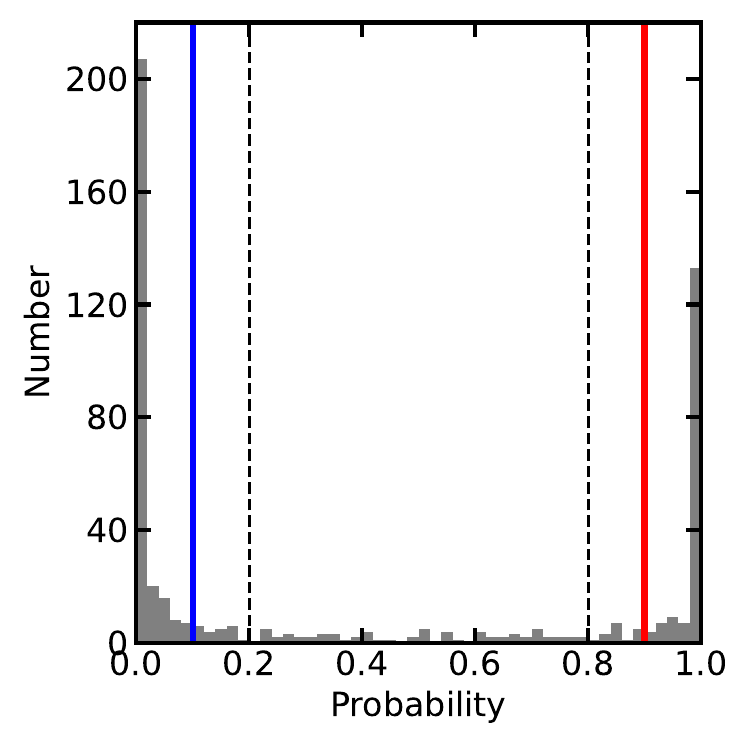}
    \caption{Confidence cuts applied to CNN prediction probabilities based on the return from the Binary Cross Entropy loss function, providing a single numeric value for each galaxy's prediction. The figure displays the probability distribution from ensemble of one-hundred models of the first redshift bin. According to our labeling system, values close to 0 signify disk class (depicted in blue), and those close to 1 are identified as spheroids (depicted in red). At this stage, we introduce a class of Irregular galaxies (depicted in two vertical dashed lines). These are galaxies that fall into the central region of the prediction probability distribution.}
    \label{fig:confident_cut}
\end{figure}

The same procedure simultaneously defines the optimal number of superclusters. This is possible because WCSS is calculated during the supercluster split, thus addressing two questions simultaneously: (1) the optimal parameter that demonstrates the lowest WCSS value for a given supercluster number, and (2) the optimal number of superclusters. Interestingly, we consistently find two to be the optimal number of superclusters. We believe this indicates that the current image resolution and metric sensitivity can split our data into only two morphological classes, supporting our claim that this methodology is suitable to separate galaxies into spheroidal and disk galaxies.

Figure \ref{fig:wcss} exhibit how the WCSS function provides an optimal supercluster split for both metrics: G in panel (a) and $\texttt{G}_\texttt{2}$ in panel (b). There is a steeper fall in the WCSS  when the number of classes change from 1 to 2, in comparison to other transitions. Our rationale is based on the fact that if splitting the data into more superclusters significantly impacts the inter-cluster variance, indicated by a steeper decrease in WCSS, then the split is justified, and the current number of superclusters better describes the data. If the slope in the WCSS function is lower, it suggests that the split only marginally improves the variability, leading to the conclusion that there may not be sufficient resolution for this separation or the metrics used are not enough to find finer subclasses.

Additionally, the order in which we supply metrics to SOMbrero can also influence results. In order to maximize the unsupervised clustering, we ensure that metrics with best discrimination between two classes will be prioritized (supplied first). At this stage, we have WCSS versus number of classes for each metric, quantifying its performance to split the data. We then order the metrics and supply to SOMbrero. We find empirically that $\texttt{G}_\texttt{2}$ and \texttt{E} are the most impactful metrics according to WCSS, and they usually are supplied as first vector entries to SOMbrero. \texttt{G} and $\texttt{M}_\texttt{2O}$ tend to score lower in terms of WCSS, and have less impact on clustering.

In summary,  the unsupervised optimization consists of two steps: First we run WSCC based optimization for each metric individually, in order to obtain WCSS value for a given combination of parameters ($n_{\rm bins}$ for \texttt{E}, \(m_{\rm tol}\) and \(p_{\rm tol}\) for \(\texttt{G}_\texttt{2}\)); Second, we run all the metrics together, but vary the parameters for the metric being optimized. By doing so, we guarantee that, in the second step, we evaluate not only the individual performance of each metric but also how it works in tandem with other metrics. We select the best parameters for \(\texttt{G}_\texttt{2}\) and \texttt{E} based on this run. Likewise, we order the metrics based on WCSS resulting from the same run. In practical terms, this strategy results in two pipeline runs. The first run focuses on optimization, selecting the best values for \(n_{\text{bins}}\) for Entropy, and $\rm m_{tol}$ and $\rm p_{tol}$ for $\texttt{G}_\texttt{2}$. The second run uses these optimized values to obtain metric values used as input for unsupervised clustering.

\subsection{Deep Learning and Classification}

Despite applying preprocessing steps to galaxy images for estimating non-parametric indices, we provide the original, unprocessed images to the Deep Learning algorithm. These are prepared by converting the initial F814W filter cutouts, originally in .fits format, to .png format using the Trilogy package from \citet{coe2012clash}. Once labels are assigned by the SOMbrero algorithm during the unsupervised stage, we move to the supervised approach with convolutional neural network (CNN) models. Compared to unsupervised labeling, once a CNN model is trained, it can be applied much faster on vast amounts of data to provide labeling. However, it is important to remember that the CNN model is only as good as the data it was trained on.

In this study, we employ a popular form of transfer learning that involves using pre-trained deep learning models. These models, trained on large-scale datasets, are repurposed for specific tasks by replacing and fine-tuning the last few layers of the network. The underlying assumption of this approach is that earlier layers of deep networks learn general, reusable features (such as edges, patterns and forms), while later layers become increasingly task-specific. Our methodology incorporates a transfer learning approach to achieve two main objectives: reducing the necessary data volume for effective training and expediting the training process by limiting the required epochs. We utilize pre-trained weights from the ImageNet dataset \citep{deng2009imagenet}. These weights, optimized to discern a plethora of features in the source domain, are aptly modified in our model to classify galaxies based on their morphological distinctions. With these pre-trained weights as a foundation, our model undergoes fine-tuning on our specific dataset during training. This retains the high-level features learned from ImageNet while adapting to the specific intricacies of our galaxy classification task. As a result, our model achieves commendable accuracy using a significantly smaller training dataset in a condensed timeframe, underscoring the efficacy of the transfer learning paradigm.

For our pipeline, we use TensorFlow as our framework. This open-source machine learning framework, developed by the Brain Team at Google \citep{abadi2016tensorflow}, offers a comprehensive ecosystem that facilitates the development and implementation of a wide range of machine learning models and complex numerical computations. TensorFlow operates through data flow graphs where nodes represent mathematical operations and edges denote multidimensional data arrays—known as tensors—that flow between these nodes. For our primary model architecture, we employ Xception, a deep learning model. ``Xception'', standing for "Extreme Inception" is a modification of the Inception architecture. Its primary innovation lies in the use of depth-wise separable convolutions, instead of the standard convolutions found in Inception. Comprising 36 convolutional layers, Xception forms the feature extraction base of our model \citep{chollet2017xception}. This architecture allows us to leverage its advanced feature extraction capabilities, improving our model's ability to classify galaxies accurately based on their morphological properties. The further steps in the pipeline are straightforward and unchanged compared to the original paper. We use the labeled data from SOMbrero as the training set, splitting it into training and validation portions. 

\subsection{CNN Confidence Thresholding}

\begin{figure*}
	\includegraphics[width=505px]{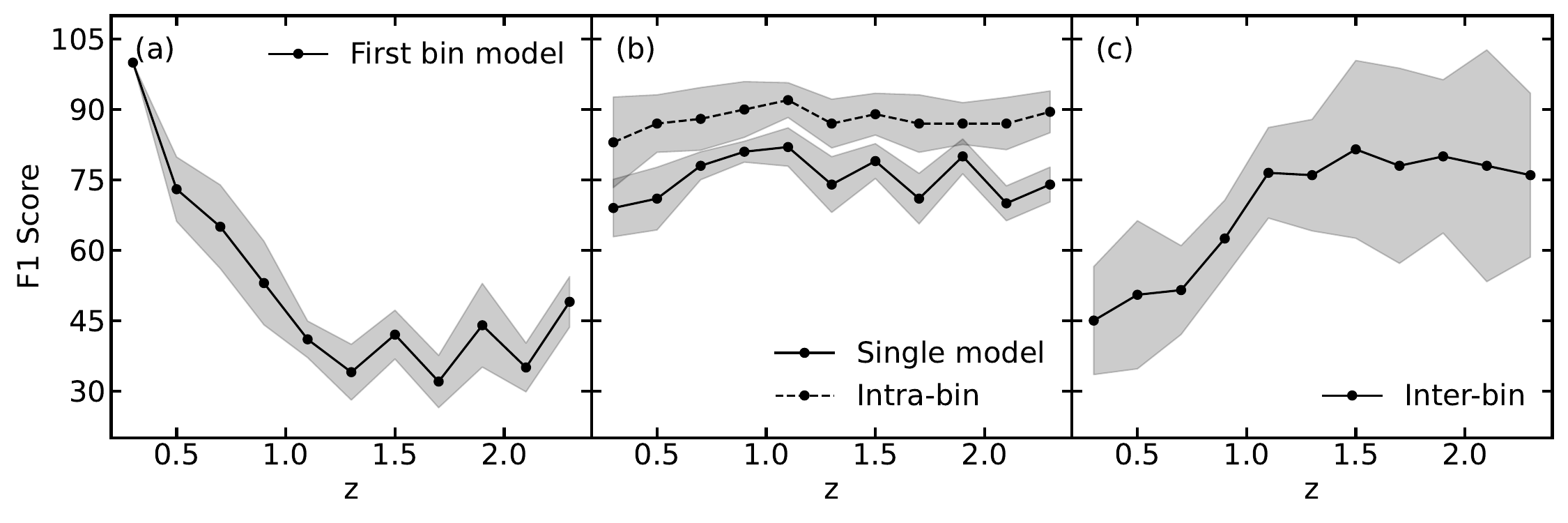}
    \caption{Showcase of model degradation across bins of z. Panel (a) demonstrates the ability of a model trained on the $0.2 < z < 0.4$ bin classifies galaxies from all other bins. The figure shows expected results, where the performance falls rapidly as the galaxies change with higher z. Panel (b) show the performance, measured by F1 Score, of the single model trained on prominent galaxies from all bins and compared with classification of in-bin models (solid black line) and performance (or inter bin variability between ensemble of 100 models) of DL inside of specific bin. In other words, the dashed line shows the consistency of model training and application on different slices of data from same bin. We note only a 75\% match of classification between the single model and models dedicated to each bin across whole z range, while dashed line is closer to 90\%. Mean values are 75\% and 87\% for single model and intra bin respectively. Lastly, panel (c) show crosmatched application of models across whole z range. We note a stark decrease of match percentage in the lover redshifts, which agrees with findings in \citet{conselice2011tumultuous}.}
    \label{fig:z_degrad}
\end{figure*}

According to our findings in K24, our labeled dataset contains approximately 3-5\% uncertainty. In other words, some galaxies were mislabeled for various reasons and we address this issue employing an ensemble training technique, which involves training multiple models on identical data but with different random selections for the training and validation datasets. By using diverse subsets of the data for each individual model in the ensemble, we introduce a level of variation that can decrease the chance of overfitting and improve the robustness of the overall predictive performance. This ensemble strategy aims to leverage the strengths of multiple models, producing more stable and reliable predictions that are less influenced by uncertainty in the training data. Specifically, we train one hundred models for each redshift bin. For each trained model we select different galaxies to be part of training and validation dataset. By this, we decrease the potential effect of misclassified galaxy have high impact on the final prediction, as all CNN probabilities will be averaged for all hundred models. This strategy not only mitigates the impact of uncertainty in the final classification but also provides us with information on the accuracy of the classifications. To further reduce the uncertainty in our training dataset (one obtained from SOMbrero clustering), we apply confidence thresholding to the final output of the deep learning classification. This methodology involves retaining data samples that exhibit a classification probability surpassing a predetermined high-confidence threshold. For instance, we use Binary Cross-Entropy as our loss function. As the final result, this provides a classification probability, a single value ranged from 0 to 1. Values closer to 0 indicate a higher likelihood of the object being a disk galaxy, and values closer to 1 indicate a higher likelihood of the object being a spheroidal. Figure \ref{fig:confident_cut} illustrates the distribution of the classification probabilities as they come from the CNN model. We note two peaks at the extremities, corresponding to the two classes, and a few intermediate values. 

Therefore, we define galaxies with probability greater than 0.9 as spheroidal, and systems with probability smaller than 0.1 as disk galaxies, as shown by the red and blue lines in Figure \ref{fig:confident_cut}. We also define a sample of irregular galaxies, characterized by a probability between 0.2 and 0.8. Nevertheless, the results touching irregular galaxies should be treated with caution, as we here assume that the inability of our method in classifying a system as disk or spheroidal is an indication that the system presents irregular features compatible with irregular galaxies. Once we apply these probability thresholds to all bins to obtain labeled samples, a small portion of galaxies that do not fall into any classifiable range are referred to as Unclassifiable and are removed from the dataset for further studies.
\section{Classifying galaxies in a wide redshift range} \label{redshift}

\begin{figure*}
	\includegraphics[width=505px]{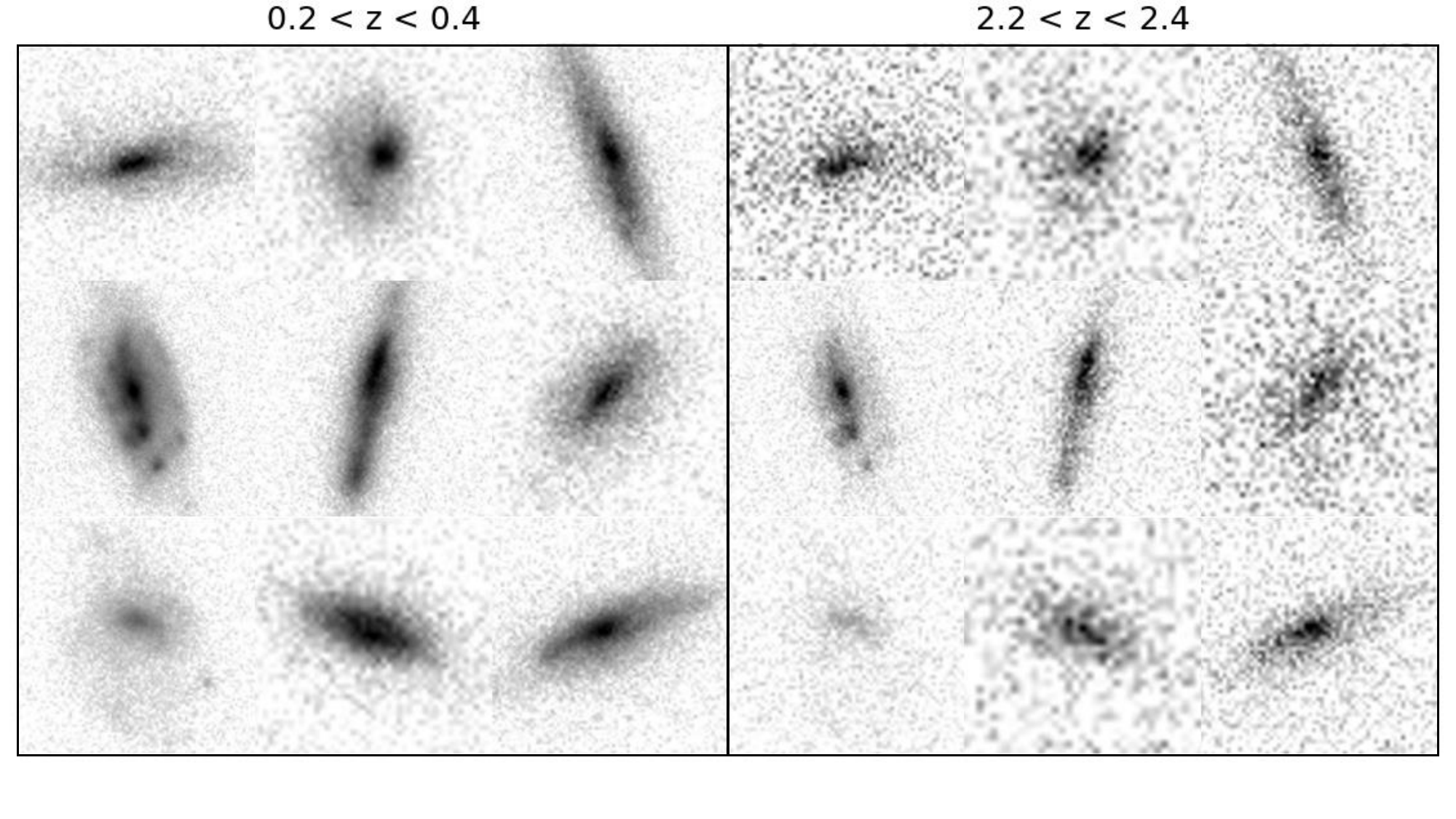}
    \caption{The figure presents several examples of the same galaxies that have been artificially redshifted. The left 3x3 mosaic displays the original selected disk galaxies, while the right shows these galaxies redshifted to a redshift of 2.3, corresponding to the last \( z \) bin. It is evident how significant the visual degradation effects are. This further confirms our concerns regarding visual classification and inspection; the majority of the features simply vanish or are no longer discernible.}
    \label{fig:degrad_mosaic}
\end{figure*}

In this section, we explore how automated methods to classify galaxy morphology rely on redshift-related effects. Particularly, previous works use the same model to a wide redshift range. For instance, the redshift range 0.2 - 2.4 comprises a time interval of 8.5 Gyr, which is considerably longer than the expected time-scale for galaxy evolution. When considering the size and mass distribution, it is noticeable that early-type galaxies are, on average, smaller than late-type galaxies at all redshifts, with a faster average size evolution for early types and a moderate evolution for late types \citep{van20143d}. Furthermore, galaxies become more concentrated at higher redshifts, suggesting that their formation may occur through mergers \citep{whitney2021galaxy}. Therefore, it is expected that using a single model in a wide redshift range may directly affect the morphological classification.

\subsection{Coarse vs Fine-grain models}\label{sub_sec:redshift_binning}

To avoid the pitfalls present on any DL modeling that use all galaxies over a wide redshift range as training dataset, we apply the method described in Section \ref{method_update} on each redshift bin shown in Figure \ref{fig:base_qtd_combined}, separately. Ultimately, this means that we have a redshift bin dedicated model trained solely on the data labeled from SOMbrero prominent clusters.

First, we test how a model trained only on low-z data performs when applied to higher redshift bins. To track accuracy when using different models to derive morphological classification, we use the F1 score -- defined as the fraction of galaxies that do not change their class between comparisons. In panel (a) of Figure \ref{fig:z_degrad}, we show the F1 score when applying the model trained using only galaxies in the first redshift bin ($0.2 \leq z < 0.4$) to all other bins. The Solid line alongside shaded regions represent the median and $\rm Q_{\sigma}$ scatter, calculated as
\begin{equation}
\label{eq:qsigma}
   \rm  Q_{\sigma} = 0.743 \times (Q_{75\%} - Q_{25\%}),
\end{equation}
where $Q_{x}$ denotes the $x^{th}$ percentile, calculated using the ensemble of one hundred models for each redshift bin. Throughout the paper, we calculate scatter in the desired quantities using equation \ref{eq:qsigma}. Exploring panel (a), we find a decreasing F1 score until $z\sim1.2$, followed by a plateau at \textbf{$\sim 40\%$}. This is an important evidence that galaxy morphology is evolving with redshift, such that there are subtleties that need to be taken into account in a more fine-grain approach, instead of the coarse-grain usually found in recent literature.

\begin{figure}
	\includegraphics[width=\columnwidth]{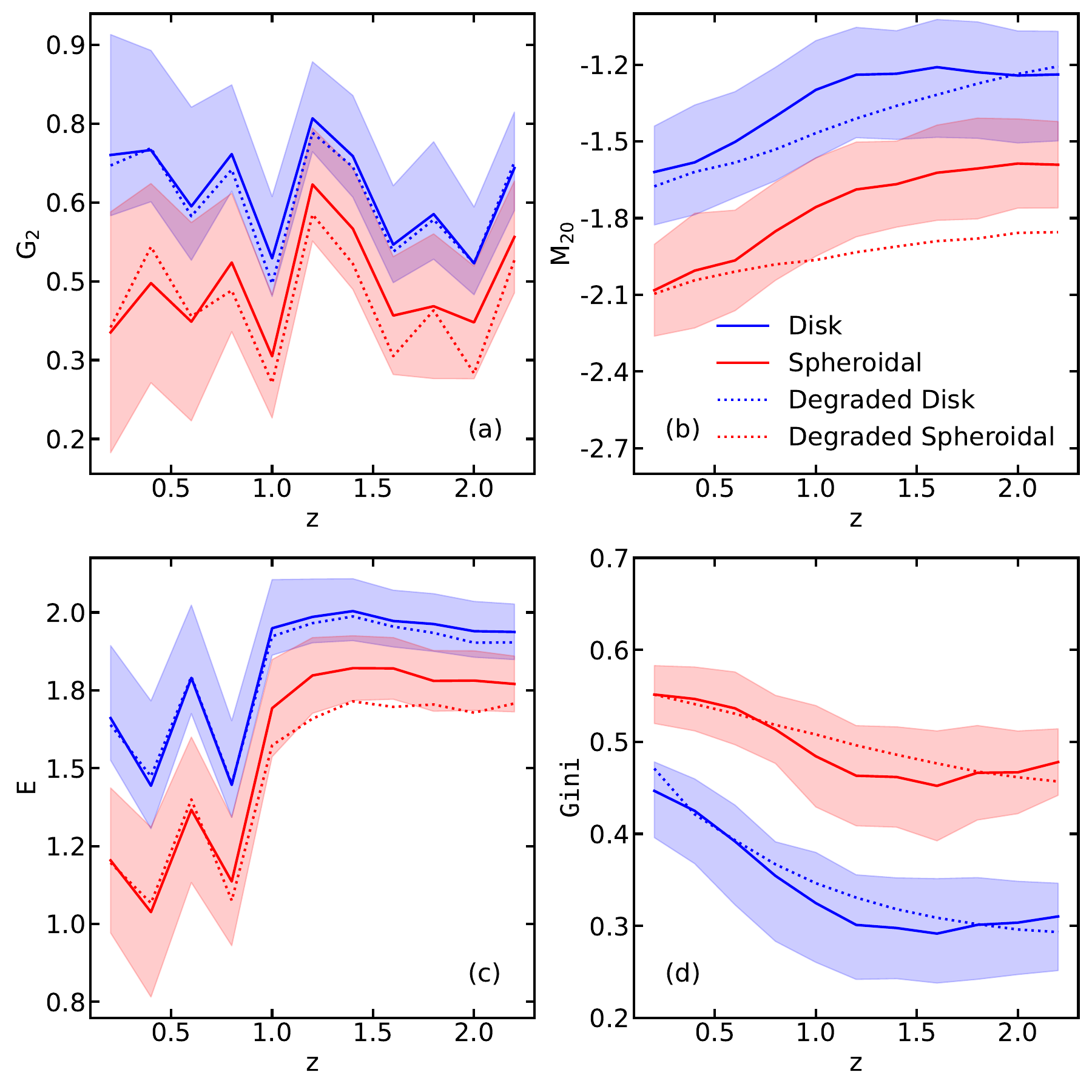}
    \caption{Metrics measured on degraded galaxies in comparison with the real disk and real spheroidal galaxies across the entire redshift range based on their metric values. Blue and Red solid lines correspond to median value of main two classes. Blue and Red dotted lines correspond to degraded set of galaxies of each class. In case of \texttt{Gini} and \(\texttt{M}_\texttt{2O}\), we note slight deviation, especially in the middle of the redshif range. In general, this behavior reinforces the confidence in our degradation procedure.}
    \label{fig:metrics_degrad}
\end{figure}

Second, we test how a single model trained on all the galaxies in the sample, irrespective of redshift, compares to bin dedicated models. In panel (b) of Figure \ref{fig:z_degrad}, we show as solid line the F1 score when comparing a single model using galaxies with the one using bin-dedicated models. We find an F1 score of  $\sim 75\%$, independent of the redshift, thus highlighting the importance of considering galaxy evolution and image degradation when assessing morphology. The change in the morphological classification of $\sim 25\%$ of the galaxies indicate that applying a single model to a wide redshift range can lead to biases due to specific morphological characteristics present at different redshifts. Nevertheless, it is noteworthy that, by attributing different models for each bins, we are also assuming that our model is capable of implicitly addressing degradation and evolution related effects. Thus, we use the one hundred ensembles in each redshift bin to investigate the variance in bin-dedicated models. The F1 score when comparing the intra-bin results is shown as dashed line in panel (b). We find an average F1 score of $\sim 90\%$, which shows consistency in our morphological classifications. Comparison between solid and dashed lines shows that, even with variations in the bin-dedicated models, the F1 score is significantly higher ($\Delta {F1 \, Score} \sim 15\%$) than adopting a single model to classify galaxies over the whole redshift domain.

Finally, in panel (c) we present a cross-match comparison between all the eleven bin-dedicated models. Namely, we classify galaxies in each redshift bin using the eleven bin-dedicated models. The comparison is done then between the classification attributed by the model dedicated to a given redshift bin, with the one assigned by the other ten models. The cross-match comparison shows two relevant trends: 1) for the $z<1$ bins, we find a decreasing F1 score, suggesting these galaxies present morphological characteristics that possibly are not in the majority of galaxies in the $z>1$ -- for instance, we find a F1 score of $45 \pm 11.5\%$ for our first redshift bin; 2) a different trend is observed for the $z\geq1$ bins, where we find an almost constant F1 score of $\sim 75 \pm 17 \%$ -- this trend indicating that the majority of models above this limit are considerably similar, despite a larger $1-\sigma$ scatter. The last trend may follow from galaxies suffering a significant morphological variation around $z \sim 1$, in agreement with previous works \citep[e.g.][]{conselice2011tumultuous}. Therefore, it is important that these changes are addressed correctly when defining a training set for a wide redshift range. Thus, we advocate a word of caution when dealing with automated algorithms for measuring galaxy morphology applied to a wide redshift range \citep[e.g.][]{hausen2020morpheus, robertson2023morpheus}. Finally, we hereon present results using bin-dedicated models to provide morphological labels to galaxies in our sample, following the thresholds shown in Figure \ref{fig:confident_cut}. 

\subsection{Degradation and evolution of disks observed at high redshifts}

When exploring higher redshift domains, numerous effects influence the observations. For instance, surface brightness dimming might obscure the fainter regions of a galaxy outskirt. The obscuration of disks, may lead to biased classifications. Furthermore, since disk galaxies show a clumpy light distribution, such that not necessarily the whole disk is obscured due to flux dimming. A region-dependent obscuration may lead to an overestimation of irregular galaxies fraction. In addition, due to the universe expansion and increasing distance, galaxies have a decreasing angular size with redshift, resulting in decreasing number of pixels available for any analysis. Both effects are significantly impactful in our work, as we extend our classification up to \( z = 2.4 \), and they can introduce additional complexities to the extraction of metrics and the performance of CNN models.

To address how degradation plus evolution affect our morphological analysis, we select all disk galaxies in the first redshift bin ($0.2 \leq z < 0.4$) and simulate how they would look at different redshifts, while keeping the same instrumental setup (exposure time, broad-band used and point-spread function), ranging from $z = 0.3$ to 2.3 in steps of 0.2. We use the Full and Efficient Redshifting of Ensembles of Nearby Galaxy Images \citep[FERENGI,][]{barden2008ferengi} code to investigate the effect of surface brightness dimming and decreasing angular size. Briefly, the code simulates the observation of galaxies originally located at redshift $z_{0}$ to a desired redshift $z'$. Mathematically, the decreasing angular size is described by
\begin{equation}
    \rm \frac{a_{z'}}{a_{z_{0}}} = \frac{d_{z_{0}}/(1+z_{0})^{2}}{d_{z'}/(1+z')^{2}},    
\end{equation}
where the left hand side is the ratio between the angular size at $\rm z'$ and $\rm z_{0}$, and $\rm d_{z'}$ and $\rm d_{z_{0}}$ are the luminosity distance at the same redshifts. Additionally, by imposing the same absolute magnitude in both redshifts, the dimming in flux follows
\begin{equation}
 \rm   \frac{f_{z'}}{f_{z_{0}}} = \left ( \frac{d_{z_{0}}}{d_{z'}} \right)^{2},    
\end{equation}
where $f_{z'}$ and $f_{z_{0}}$ are the flux observed at each redshift. Notably, the code also allows for a parametrization of the luminosity evolution of galaxies. This is particularly relevant to our work, as we touch redshifts consistent with the peak in cosmic star formation density $z \sim 2-3$ \citep{2014ARA&A..52..415M}. We  use  an absolute magnitude correction, following the linear relation $\rm M_{z'} = M_{z_{0}} - (1 \times z')$, where $\rm M_{z'}$ and $\rm M_{z_{0}}$ are the absolute magnitude at $\rm z'$ and $\rm z_{0}$ \citep{barden2008ferengi}, respectively. We fix the evolution term as -1.0 simply as a way to show the impact of it. However, establishing a more accurate functional form for evolution is beyond the scope of this investigation, especially considering its uncertainty \citep{1961ApJ...134..916S, 1997A&AS..122..399P, 1998ApJ...507..497C, 2001Sci...293.1273A}. For completeness, we performed the same exercise for spheroidal galaxies and present the degradation results alongside.

We calculate the \texttt{MEGG} metrics for the redshifted galaxies to ensure consistency with real galaxies from corresponding redshift bins. Figure \ref{fig:metrics_degrad} illustrates the metrics derived from redshifted galaxies compared with those from real galaxies in corresponding bins. It is noteworthy that, while we do note a relevant metric variations with redshift, the simulated galaxies show consistent trends with real galaxies. More importantly, we note larger variations in $\texttt{M}_\texttt{2O}$ and \texttt{G} index, in comparison to $\texttt{G}_\texttt{2}$ and \texttt{E}. We thus compare the relevance of each metric for the definition of prominent clusters, for which we find that, not by chance, the last two are the most relevant.

\begin{figure}
	\includegraphics[width=\columnwidth]{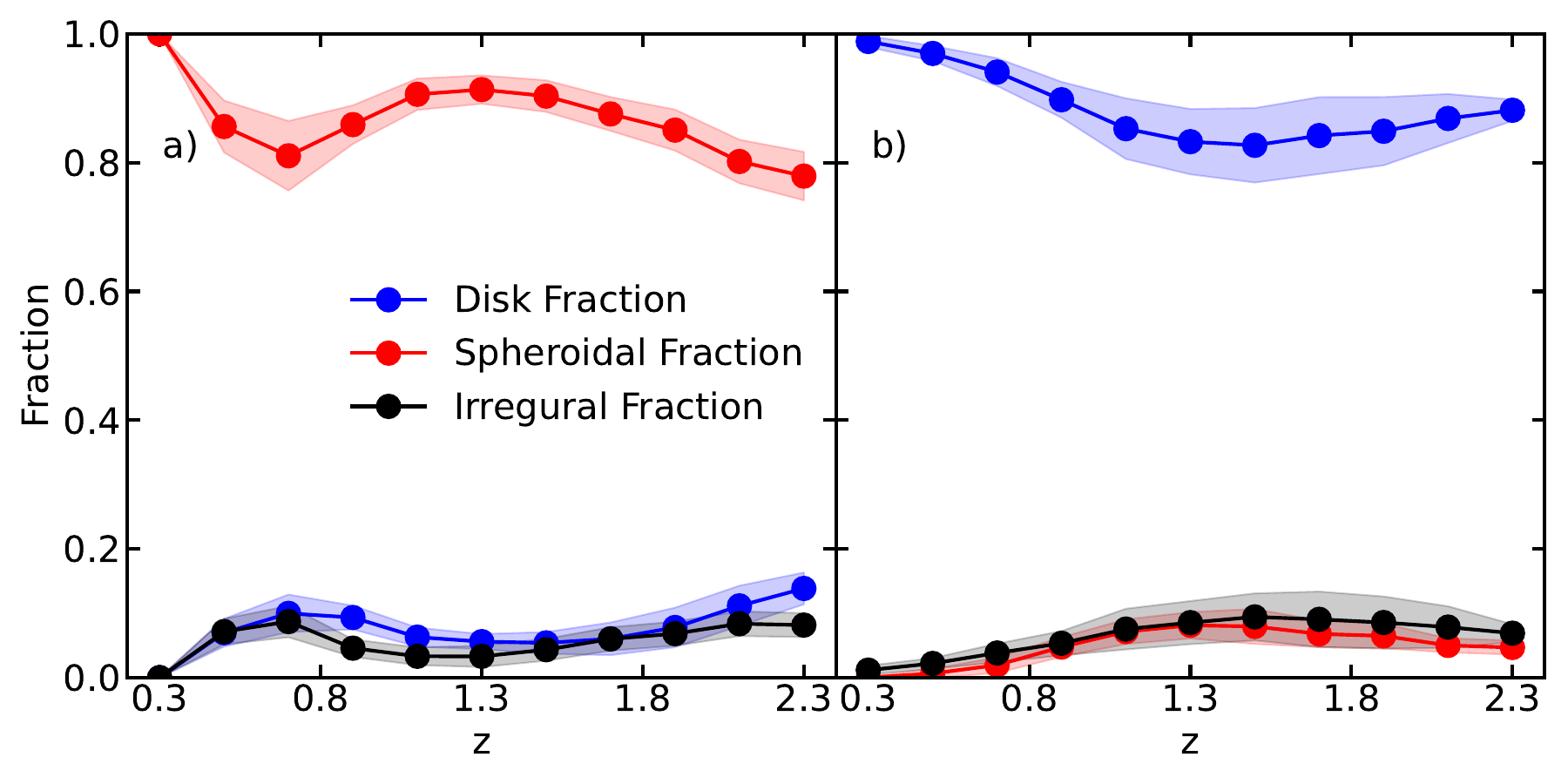}
    \caption{Degradation results for two main classes; spheroidal in panel (a) and disk in panel (b). Each panel contains fractions of disk, spheroidal and irregular galaxies (from the simulated sample only) across our redshift range. It can be noted that that both classes showcase a slight decrease at the beginning and then stabilizes around 85\%, while similar increase of Irregular and spheroidal galaxies is noted. This indicates that, while some galaxies are misclassified as these classes due to redshift effects, this number hovers around 15\%. Mean fraction for spheroidal degradation is 92\% and for disk is 95\% which are expected in prism of comparison of overall method accuracy achieved in K24.}
    \label{fig:degrad_frac}
\end{figure}

Using a set of artificially redshifted galaxies, we investigate how degradation and luminosity evolution effects may lead to misclassification in galaxy morphology. We take all galaxies in the redshift bin $0.2 < z < 0.4$ classified as spheroidal or disks and artificially degrade them to simulate observations at higher redshift bins from 0.4 to 2.4. We then classify the degraded galaxies using models dedicated to each specific redshift bin, constructed from real data—that is, we use models trained on data from Figure \ref{fig:base_qtd_combined}. Figure \ref{fig:degrad_frac} presents the results for the two separate classes: spheroidal galaxies (in red, panel a) and disks (in blue, panel b). After performing inference with our dedicated models, we measure how many galaxies maintain their classes, defining the fraction displayed in the figure. Notably, both classes demonstrate similar behavior, which can be largely attributed to the uncertainty of the method, resulting in mean fractions of 92\% for spheroidal and 95\% for disks. As presented in K24, we expect an accuracy of approximately 90\%. The observed decrease of about 15\% on average reinforces how spheroidal and disk galaxies can be misclassified simply due to degradation.


\section{Results and Discussion}\label{sec:results}

While in the local universe the morphology of galaxies can be easily classified by visual inspection -- therefore defining a reliable training set, at higher redshifts flux dimming, observational limitations and the presence of irregular features input a higher degree of subjectivity in visual inspection. Our work is focused on providing a reliable classification that do not depend on training sets built by visual classification. Next, we present our main results derived by applying the method described in Section \ref{method_update} to CANDELS galaxies.

\begin{figure}
	\includegraphics[width=\columnwidth]{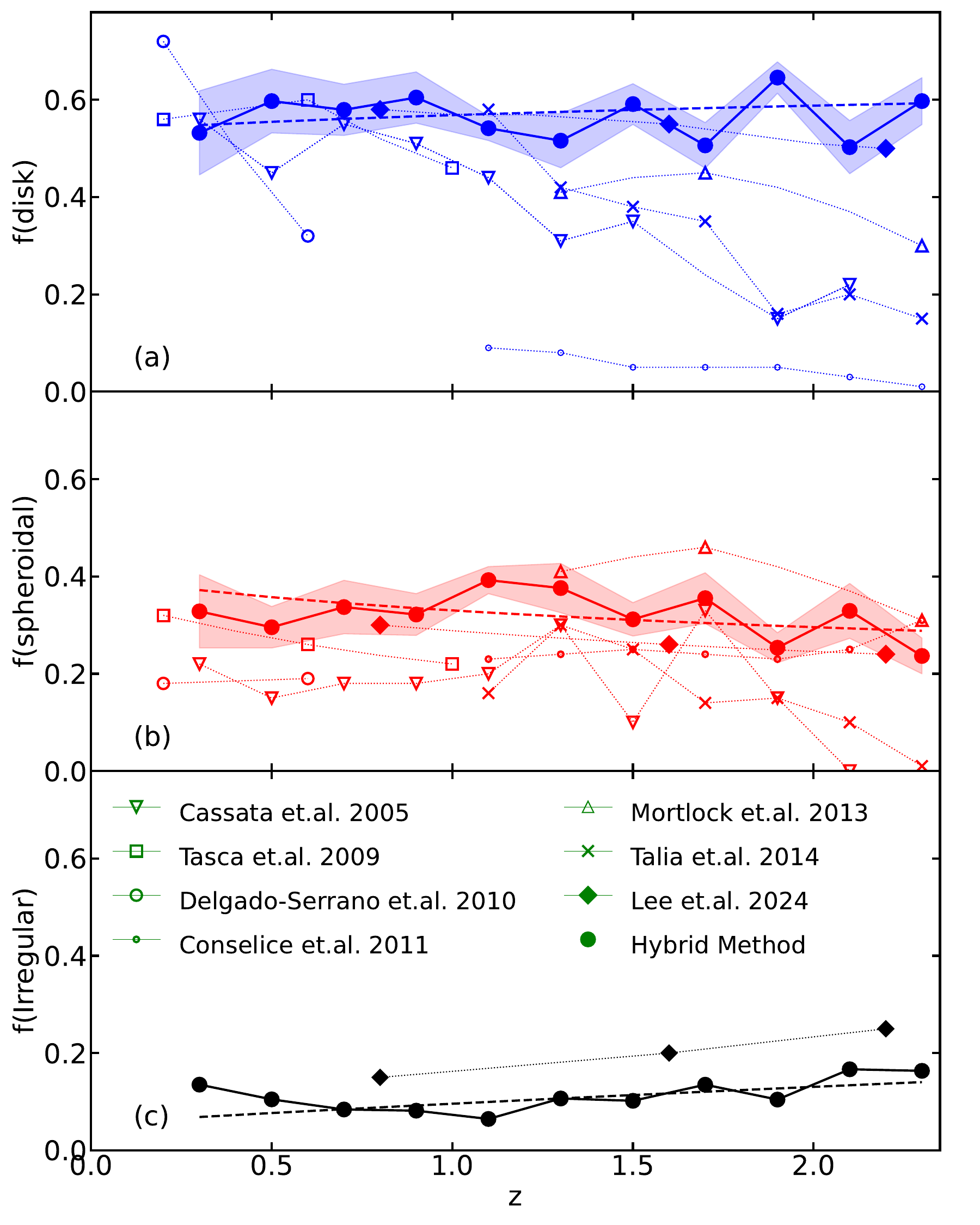}
    \caption{Comparison of the fractions obtained in this research with ones presented in the literature over last decade. These fractions are plotted across the redshifts for disk, spheroidal and Irregular galaxies. It is notable that we we tend to agree with most of the findings, for both main classes, reinforcing the robustness of the method. The dashed lines show the result of fitting the data to equation \ref{eq:fraction}. To produce the error bars for our ratios we used performance of each of our one hundred ensemble models to calculate $Q_{\sigma}$ and median. Tabular fractions can be found in Table \ref{table:main_frac}.}
    \label{fig:frac_main}
\end{figure}

A particular advantage of analyzing a wide redshift range is to probe how galaxies evolve over a large look back time interval. In particular for morphology, the fraction of disk, spheroidal and irregular galaxies have been the focus in the recent literature due to its implication to galaxy formation and evolution theory \citep{cassata2005evolution,tasca2009zcosmos,conselice2011tumultuous,mortlock2013redshift, talia2014listening, kartaltepe2015candels, lee2024morphology}. Therefore, we show in Figure~\ref{fig:frac_main}, respectively from top to bottom panels, the fraction of disk (blue), spheroidal (red) and irregular (black) galaxies according to our method (solid line), and according to previous works in the literature (dotted lines). Shaded areas around the solid line represents the scatter in our estimate calculated using one hundred ensembles. Empty symbols represent works that make use of visual classification, whereas filled symbols show the results from works that adopt objective -- i.e. does not depend on visual inspection -- criteria to define morphology. In the case irregular galaxies fraction, we only include the results from \citet{lee2024morphology}, as the definition of irregular galaxies can be somewhat ambiguous and this is the only work in which the adopted definition is consistent with ours, i.e. follow from an objective criteria. 

Examining Figure \ref{fig:frac_main}, we find that the fractions are highly dependent on the adopted method. Significant differences are observed in the $z > 1$ region, especially for disks, when comparing our results with those from visual inspection. This discrepancy could be even larger considering the effect shown in Figure~\ref{fig:degrad_frac}. Conversely, despite the use of different data and methodologies, our findings closely align with those in \citet{lee2024morphology}, which also avoids visual inspection in morphological classification. We therefore suggest that visual inspection, whether directly or for creating training sets, should be avoided to prevent bias from the effects discussed in Section~\ref{redshift}. Notably, while our classification relies on a single band (F814W), \cite{lee2024morphology} vary the observed band with redshift to always probe the same rest-frame wavelength for all galaxies, irrespective of redshift. Thus, the similarity in the results suggests minimal variations in the non-parametric indexes across different wavelengths, even though the stellar population evolves significantly over the redshift range.

Comparison between disk, spheroidal and irregular fractions over the interval $0.2 < z < 2.4$, shows an almost constant fraction for all three, irrespective of redshift. Disks comprises roughly $\sim 60\%$ of galaxies, whereas spheroidal and irregular galaxies represent $\sim 30\%$ and $\sim 10\%$, respectively. We quantify variations in the fraction as a function of redshift using the following relation:
\begin{equation}
\label{eq:fraction}
\rm    F({\rm class}) = C_{0} \times (1+z)^{m_{\rm class}},
\end{equation}
where ``class'' stands for disks, spheroids or peculiars. We show in Table~\ref{tab:fit_results} the fitting results for the curves shown in Figure~\ref{fig:frac_main}.

Regarding disk fraction, our fitting results ($m_{\rm disk} = 0.08$) reinforces an almost flat fraction as a function of redshift. 
Notably, this is in disagreement with most observations characterizing galaxies morphology beyond $z > 1$ \citep{cassata2005evolution, mortlock2013redshift, talia2014listening, delgado2010hubble, tasca2009zcosmos}, that suggests a decreasing fraction of disks towards higher redshifts. We suggest that this disagreement follows from limitations of visual-inspection based classification, as it gets increasingly harder to detect disks for increasing redshift.

With respect to irregular systems, we find a slight ($\sim$10\%) increase for the $z > 1$ region. Our fitting results suggest that the fraction of irregular galaxies increases with redshift, doubling its initial value after a $\Delta z \sim 1.57$. Despite, previous works also find an increasing fraction of irregular galaxies with redshift, our increase is considerably less steeper than previously reported. Altogether, we suggest that, when probing faint galaxies at high redshifts, disks may still have notable visible regions due to its clumpiness that, although not sufficient to characterize the galaxy as a disk, may induce an excess of irregular systems when classified by eye, simply because dimming makes the disk not visible but the knots in the disk characteristics of HII regions are identified by the eye.

\begin{figure*}
	\includegraphics[width=505px]{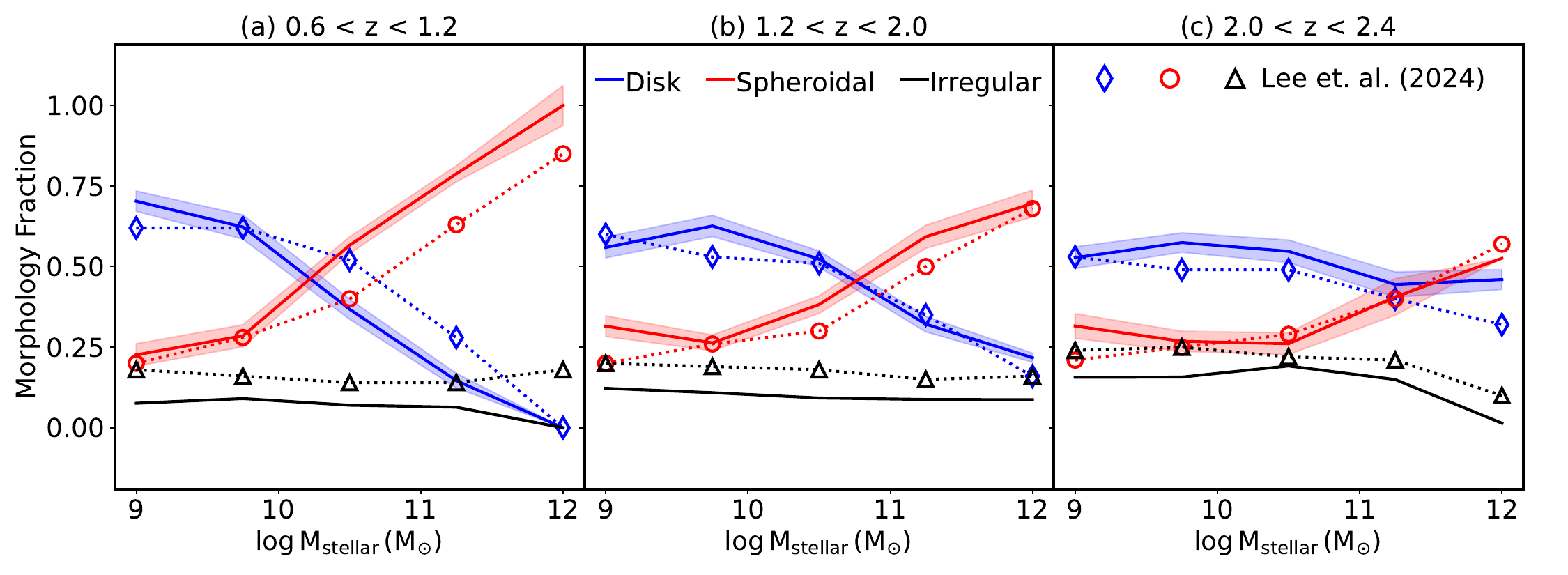}
    \caption{Companion figure to Figure \ref{fig:frac_main}. Morphology fractions as a function of Stellar mass in three segments of redshift. Solid lines correspond to Hybrid Method results while dotted lines with markers correspond to the results of \citet{lee2024morphology}. It is noteworthy the compatibility of the results given that two researches used different datasets (HST and JWST) and different methods. The only point of intersection is that both methods are Unsupervised-Supervised and data driven. To produce the error bars for our ratios we used performance of each of our one hundred ensemble models to calculate $Q_{\sigma}$ and median. Important note that errors in this Figure are compatible with the error presented in the Figure \ref{fig:frac_main}. Tabular fractions can be found in Table \ref{table:super_lee_fractions1}}
    \label{fig:super_lee}
\end{figure*}

Exploring results shown in Table~\ref{tab:fit_results}, we find that, contrary to the trend observed for disks and irregulars, the fraction of spheroids decreases with redshift. Quantitatively, it decreases from $\sim 36\%$ to $\sim 20\%$ between $z = 0.2$ and $z = 2.4$, respectively. Therefore, we suggest that the formation of the majority of spheroidal galaxies may follow from evolution related mechanisms, such as mergers \citep{2009ApJ...697.1369B, 2014MNRAS.444.1475M, 2017MNRAS.467.3083R, 2019MNRAS.489.4679J}, especially for massive systems, and AGN/stellar feedback \citep{2013MNRAS.432.1989S, 2016ApJ...820..131E, 2024MNRAS.532.2558Z}, particularly relevant for less massive systems, both previously pointed as triggers for morphological transition \citep[e.g.][]{2022MNRAS.509..567S, 2024MNRAS.532..982S}. Still, an analysis of galaxy evolution as a function of redshift will be presented in a future work (Sampaio et al. in prep).

Furthermore, galaxy evolution, including morphological evolution, depends on stellar mass. To assess the evolution of galaxy morphology as a function of stellar mass and redshift, we define three different redshift bins: $0.6 \leq z < 1.2$, $1.2 \leq z < 2.0$, and $2.0 \leq z < 2.4$, which are selected for consistency with \citet{lee2024morphology}. Figure~\ref{fig:super_lee} shows the fraction of spheroidal (red curve), disk (blue), and irregular (black) galaxies as a function of stellar mass for the three bins mentioned above. The shaded areas denote the $1-\sigma$ scatter derived from the one-hundred ensembles (See Section \ref{redshift}). The dotted lines represent the results from \citet{lee2024morphology}. 

This comparison shows a striking similarity between our results and those of \citet{lee2024morphology}, reinforcing the agreement when using objective methods to characterize morphology. Exploring the relation shown for irregular galaxies, we find an approximately constant fraction, irrespective of stellar mass and redshift. The only notable variation is in panel (c), which shows a slight decrease in the fraction of irregular galaxies for stellar masses above $10^{10.5}{\rm M}_{\odot}$. Moreover, the most significant result comes from comparing disk and spheroidal galaxies. The fractions of disks and spheroids exhibit nearly opposite behaviors. In panel (a), the fraction of disks decreases by $\sim$66\% over the entire stellar mass range, while the fraction of spheroids increases similarly by $\sim$74\%. The same pattern is seen in panel (b), where the fraction of disks decreases by $\sim$32\%, while the fraction of spheroids increases by $\sim$34\%. In panel (c), the fraction of disks remains constant up to stellar masses of $10^{10.5}{\rm M}_{\odot}$ and then decreases by $\sim$7\%. The fraction of spheroids in this highest redshift bin shows an opposite trend compared to disks, constant up to stellar masses of $10^{10.5}{\rm M}_{\odot}$ and then increasing by $\sim$24\%. 

\begin{table}
\centering
\caption{Results for the fitting of equation \ref{eq:fraction} to the curves shown in Figure~\ref{fig:frac_main}. Column (1) shows the class considered. Column (2) and (3) show the constant and exponent, alongside related uncertainties, for the fitting of equation \ref{eq:fraction} to the observed fractions.}
\label{tab:fit_results}
\resizebox{0.8\columnwidth}{!}{%
\begin{tabular}{c|c|c}
\hline
Class    & $C_{0}$         & $m_{\rm class}$  \\ \hline
Disk     & $0.54 \pm 0.05$ & $0.08 \pm 0.03$  \\ \hline
Spheroidal & $0.39 \pm 0.07$ & $-0.26 \pm 0.04$ \\ \hline
Irregular & $0.06 \pm 0.04$ & $0.73 \pm 0.05$  \\ \hline
\end{tabular}%
}
\end{table}

It is important to emphasize that these trends are almost identical to those observed by \citet{lee2024morphology}. The combined trends for disks and spheroids suggest that low-mass galaxies at low redshift are predominantly disks, in agreement with the downsizing scenario, assuming that most star formation occurs in disk galaxies. An important result from this investigation, also supported by \citet{lee2024morphology}, is that morphological transformations over this redshift interval ($\sim 5$ Gyr) depends on the stellar mass and redshift
in a very complex manner. This may reflect the interplay of the several parameters defining the merger of two galaxies (e.g. \citep{2015MNRAS.449...49R}).  it is expected that mergers are a fundamental step in the mass assembly of galaxies. Different works suggest that the merging of two disks is an efficient way to generate spheroidal galaxies \citep{1972ApJ...178..623T, 1992ARA&A..30..705B, 2003ApJ...597L.117K, 2005ApJ...622L...9S, 2006ApJS..163....1H} and the decreasing fraction of high-mass disks with redshift, alongside the increase of massive spheroidal galaxies, may indicate that galaxies evolve through mergers but the fractions we measure show a more complex picture.

\section{Summary}\label{section_conclusions}

In this work, we classify 13,988 galaxies from the CANDELS fields, Disks, Spheroids, and Irregulars, using a hybrid unsupervised-supervised method founded in non-parametric morphological metrics. To guarantee robustness, we apply $\rm H \leq 24$ and $\rm M_{stellar} \geq 10^{9}M_{\odot}$ cuts to ensure reliability in the metrics estimation and avoid the inclusion of dwarf galaxies in our sample, respectively. In contrast to the commonly adopted \texttt{CAS} system, we use the \texttt{EGG} system -- which was previously investigated and shows a better performance in discriminating spheroids and disks. We further include the \texttt{$\rm M_{20}$} metric, as it also shows good performance, therefore defining the \texttt{MEGG} system. 

Differently from previous works, our method does not use visual inspection in any step. We first use the metrics as input to a Self-Organizing Map algorithm to define two prominent samples of disk and spheroidal galaxies. Then, this sample is used to train a CNN algorithm to provide final morphological classification. Furthermore, we provide a in-detail analysis of the limitations and biases that may occur in morphological classification when applying CNN models created over a wide range of redshift in comparison to bin-dedicated models. In this regard, one of our main conclusion is that bin-dedicated models show a better accuracy in comparison to a model built using data from the whole redshift range (Figure~\ref{fig:z_degrad}.) We argue that this follows from the whole redshift range covering a look-back time much larger than the expected for galaxy evolution related processes, thus increasing the inaccuracy of automated methods. Thus, we separate galaxies in redshift bins from 0.2 to 2.4, in steps of 0.2, with the step width selected to guarantee at least 500 objects in each bin.

Our results provide a new perspective on morphological evolution as a function of redshift, and support that methods relying in visual inspection can introduce relevant biases. Here we highlight the main findings of this work:
\begin{itemize}
    \item By using the FERENGI code, we show that up to 17\% of disk galaxies (Figure \ref{fig:degrad_frac}) may be misclassified due to flux dimming, decreasing size and luminosity evolution related effects at redshifts $z > 1$. This should be taken into account when analyzing the fraction of galaxies with different morphology, as it may introduce several bias in visual classification;

    \item Contrary to studies suggesting a decreasing fraction of disks beyond $z>1$, our morphological analysis results in disk and spheroidal fractions approximately constant irrespective of redshift, in excellent agreement with \citet{lee2024morphology}. Quantitatively, the fraction of disks is significantly higher ($\sim 60\%$) than the fraction of spheroidal galaxies ($\sim 30\%$) over the redshift interval we study. Comparison between disk and spheroidal fractions suggest a ratio of approximately 1:2; 

    \item Comparison between fractions as a function of both redshift and stellar mass (Figure~\ref{fig:super_lee}) reveals a distinct trend for disk and spheroidal galaxies. We find a decreasing fraction of massive disk galaxies with redshift, in comparison to an increasing fraction of massive spheroidal systems. We suggest this complementary trends as an indication of the role of mergers in galaxy evolution. Namely, different works point out that mergers are an efficient way to generate a spheroidal galaxies. However, the determined fractions display the intricacies of the merger history, meaning that this process is probably more complex than a simple transformation.
    
\end{itemize}
\section{Acknowledgements}
This work was financed in part by the Coordenação de Aperfeiçoamento de Pessoal de Nível Superior – Brazil (CAPES) – Finance Code 001. VMS and RRdC acknowledge the support from FAPESP through the grants 2020/15245-2 and 2020/16243-3, respectively. CJC acknowledges support from the ERC Advanced Investigator Grant EPOCHS (788113).
\section*{Data Availability}
The data that support the findings of this study are available from the corresponding author, IK, upon reasonable request.
 


\bibliographystyle{mnras}
\bibliography{ref} 

\begin{thebibliography}{}
\makeatletter
\relax
\def\mn@urlcharsother{\let\do\@makeother \do\$\do\&\do\#\do\^\do\_\do\%\do\~}
\def\mn@doi{\begingroup\mn@urlcharsother \@ifnextchar [ {\mn@doi@} {\mn@doi@[]}}
\def\mn@doi@[#1]#2{\def\@tempa{#1}\ifx\@tempa\@empty \href {http://dx.doi.org/#2} {doi:#2}\else \href {http://dx.doi.org/#2} {#1}\fi \endgroup}
\def\mn@eprint#1#2{\mn@eprint@#1:#2::\@nil}
\def\mn@eprint@arXiv#1{\href {http://arxiv.org/abs/#1} {{\tt arXiv:#1}}}
\def\mn@eprint@dblp#1{\href {http://dblp.uni-trier.de/rec/bibtex/#1.xml} {dblp:#1}}
\def\mn@eprint@#1:#2:#3:#4\@nil{\def\@tempa {#1}\def\@tempb {#2}\def\@tempc {#3}\ifx \@tempc \@empty \let \@tempc \@tempb \let \@tempb \@tempa \fi \ifx \@tempb \@empty \def\@tempb {arXiv}\fi \@ifundefined {mn@eprint@\@tempb}{\@tempb:\@tempc}{\expandafter \expandafter \csname mn@eprint@\@tempb\endcsname \expandafter{\@tempc}}}

\bibitem[\protect\citeauthoryear{Abadi et~al.,}{Abadi et~al.}{2016}]{abadi2016tensorflow}
Abadi M.,  et~al., 2016, in 12th USENIX symposium on operating systems design and implementation (OSDI 16). pp 265--283

\bibitem[\protect\citeauthoryear{{Abraham} \& {van den Bergh}}{{Abraham} \& {van den Bergh}}{2001}]{2001Sci...293.1273A}
{Abraham} R.~G.,  {van den Bergh} S.,  2001, \mn@doi [Science] {10.1126/science.1060855}, \href {https://ui.adsabs.harvard.edu/abs/2001Sci...293.1273A} {293, 1273}

\bibitem[\protect\citeauthoryear{Abraham et~al.}{Abraham et~al.}{2003}]{abraham2003}
Abraham R.~G.,  et~al., 2003, The Astrophysical Journal, 588, 218

\bibitem[\protect\citeauthoryear{Barchi et~al.,}{Barchi et~al.}{2020}]{barchi2020machine}
Barchi P.~H.,  et~al., 2020, Astronomy and Computing, 30, 100334

\bibitem[\protect\citeauthoryear{Barden, Jahnke  \& H{\"a}u{\ss}ler}{Barden et~al.}{2008}]{barden2008ferengi}
Barden M.,  Jahnke K.,   H{\"a}u{\ss}ler B.,  2008, The Astrophysical Journal Supplement Series, 175, 105

\bibitem[\protect\citeauthoryear{{Barnes} \& {Hernquist}}{{Barnes} \& {Hernquist}}{1992}]{1992ARA&A..30..705B}
{Barnes} J.~E.,  {Hernquist} L.,  1992, \mn@doi [\araa] {10.1146/annurev.aa.30.090192.003421}, \href {https://ui.adsabs.harvard.edu/abs/1992ARA&A..30..705B} {30, 705}

\bibitem[\protect\citeauthoryear{Barro et~al.,}{Barro et~al.}{2019}]{barro2019candels}
Barro G.,  et~al., 2019, The Astrophysical Journal Supplement Series, 243, 22

\bibitem[\protect\citeauthoryear{Bello, Zoph, Vaswani, Shlens  \& Le}{Bello et~al.}{2019}]{Bello_2019_ICCV}
Bello I.,  Zoph B.,  Vaswani A.,  Shlens J.,   Le Q.~V.,  2019, in Proceedings of the IEEE/CVF International Conference on Computer Vision (ICCV).

\bibitem[\protect\citeauthoryear{Bishop}{Bishop}{2007}]{bishop2007pattern}
Bishop C.~M.,  2007, Pattern recognition and machine learning (information science and statistics)

\bibitem[\protect\citeauthoryear{{Bruce} et~al.,}{{Bruce} et~al.}{2012}]{2012MNRAS.427.1666B}
{Bruce} V.~A.,  et~al., 2012, \mn@doi [\mnras] {10.1111/j.1365-2966.2012.22087.x}, \href {https://ui.adsabs.harvard.edu/abs/2012MNRAS.427.1666B} {427, 1666}

\bibitem[\protect\citeauthoryear{{Bruzual}}{{Bruzual}}{2007}]{2007ASPC..374..303B}
{Bruzual} G.,  2007, in {Vallenari} A.,  {Tantalo} R.,  {Portinari} L.,   {Moretti} A.,  eds,  Astronomical Society of the Pacific Conference Series Vol. 374, From Stars to Galaxies: Building the Pieces to Build Up the Universe. p.~303 (\mn@eprint {arXiv} {astro-ph/0702091}), \mn@doi{10.48550/arXiv.astro-ph/0702091}

\bibitem[\protect\citeauthoryear{{Bruzual} \& {Charlot}}{{Bruzual} \& {Charlot}}{2003}]{2003MNRAS.344.1000B}
{Bruzual} G.,  {Charlot} S.,  2003, \mn@doi [\mnras] {10.1046/j.1365-8711.2003.06897.x}, \href {https://ui.adsabs.harvard.edu/abs/2003MNRAS.344.1000B} {344, 1000}

\bibitem[\protect\citeauthoryear{{Bundy}, {Fukugita}, {Ellis}, {Targett}, {Belli}  \& {Kodama}}{{Bundy} et~al.}{2009}]{2009ApJ...697.1369B}
{Bundy} K.,  {Fukugita} M.,  {Ellis} R.~S.,  {Targett} T.~A.,  {Belli} S.,   {Kodama} T.,  2009, \mn@doi [\apj] {10.1088/0004-637X/697/2/1369}, \href {https://ui.adsabs.harvard.edu/abs/2009ApJ...697.1369B} {697, 1369}

\bibitem[\protect\citeauthoryear{Cassata et~al.,}{Cassata et~al.}{2005}]{cassata2005evolution}
Cassata P.,  et~al., 2005, Monthly Notices of the Royal Astronomical Society, 357, 903

\bibitem[\protect\citeauthoryear{Cheng et~al.,}{Cheng et~al.}{2021}]{cheng2021galaxy}
Cheng T.-Y.,  et~al., 2021, Monthly Notices of the Royal Astronomical Society, 507, 4425

\bibitem[\protect\citeauthoryear{Cheng et~al.,}{Cheng et~al.}{2023}]{cheng2023lessons}
Cheng T.-Y.,  et~al., 2023, Monthly Notices of the Royal Astronomical Society, 518, 2794

\bibitem[\protect\citeauthoryear{Chollet}{Chollet}{2017}]{chollet2017xception}
Chollet F.,  2017, in Proceedings of the IEEE conference on computer vision and pattern recognition. pp 1251--1258

\bibitem[\protect\citeauthoryear{{Cimatti}, {Fraternali}  \& {Nipoti}}{{Cimatti} et~al.}{2019}]{2019igfe.book.....C}
{Cimatti} A.,  {Fraternali} F.,   {Nipoti} C.,  2019, {Introduction to Galaxy Formation and Evolution: From Primordial Gas to Present-Day Galaxies}.
Cambridge University Press

\bibitem[\protect\citeauthoryear{Coe et~al.,}{Coe et~al.}{2012}]{coe2012clash}
Coe D.,  et~al., 2012, The Astrophysical Journal, 757, 22

\bibitem[\protect\citeauthoryear{{Conselice}}{{Conselice}}{2014}]{2014ARA&A..52..291C}
{Conselice} C.~J.,  2014, \mn@doi [\araa] {10.1146/annurev-astro-081913-040037}, \href {https://ui.adsabs.harvard.edu/abs/2014ARA&A..52..291C} {52, 291}

\bibitem[\protect\citeauthoryear{Conselice, Bluck, Ravindranath, Mortlock, Koekemoer, Buitrago, Gr{\"u}tzbauch  \& Penny}{Conselice et~al.}{2011}]{conselice2011tumultuous}
Conselice C.,  Bluck A.,  Ravindranath S.,  Mortlock A.,  Koekemoer A.,  Buitrago F.,  Gr{\"u}tzbauch R.,   Penny S.,  2011, Monthly Notices of the Royal Astronomical Society, 417, 2770

\bibitem[\protect\citeauthoryear{{Contardo}, {Steinmetz}  \& {Fritze-von Alvensleben}}{{Contardo} et~al.}{1998}]{1998ApJ...507..497C}
{Contardo} G.,  {Steinmetz} M.,   {Fritze-von Alvensleben} U.,  1998, \mn@doi [\apj] {10.1086/306350}, \href {https://ui.adsabs.harvard.edu/abs/1998ApJ...507..497C} {507, 497}

\bibitem[\protect\citeauthoryear{Cottrell \& Letr{\'e}my}{Cottrell \& Letr{\'e}my}{2005}]{cottrell2005use}
Cottrell M.,  Letr{\'e}my P.,  2005, Neurocomputing, 63, 193

\bibitem[\protect\citeauthoryear{Cottrell, Ibbou  \& Letr{\'e}my}{Cottrell et~al.}{2004}]{cottrell2004som}
Cottrell M.,  Ibbou S.,   Letr{\'e}my P.,  2004, Neural Networks, 17, 1149

\bibitem[\protect\citeauthoryear{Dai, Qi, Xiong, Li, Zhang, Hu  \& Wei}{Dai et~al.}{2017}]{Dai_2017_ICCV}
Dai J.,  Qi H.,  Xiong Y.,  Li Y.,  Zhang G.,  Hu H.,   Wei Y.,  2017, in Proceedings of the IEEE International Conference on Computer Vision (ICCV).

\bibitem[\protect\citeauthoryear{Delgado-Serrano, Hammer, Yang, Puech, Flores  \& Rodrigues}{Delgado-Serrano et~al.}{2010}]{delgado2010hubble}
Delgado-Serrano R.,  Hammer F.,  Yang Y.,  Puech M.,  Flores H.,   Rodrigues M.,  2010, Astronomy \& Astrophysics, 509, A78

\bibitem[\protect\citeauthoryear{Deng, Dong, Socher, Li, Li  \& Fei-Fei}{Deng et~al.}{2009}]{deng2009imagenet}
Deng J.,  Dong W.,  Socher R.,  Li L.-J.,  Li K.,   Fei-Fei L.,  2009, in 2009 IEEE conference on computer vision and pattern recognition. pp 248--255

\bibitem[\protect\citeauthoryear{Dom{\'\i}nguez~S{\'a}nchez, Huertas-Company, Bernardi, Tuccillo  \& Fischer}{Dom{\'\i}nguez~S{\'a}nchez et~al.}{2018}]{dominguez2018improving}
Dom{\'\i}nguez~S{\'a}nchez H.,  Huertas-Company M.,  Bernardi M.,  Tuccillo D.,   Fischer J.,  2018, Monthly Notices of the Royal Astronomical Society, 476, 3661

\bibitem[\protect\citeauthoryear{{Dressler}}{{Dressler}}{1980}]{Dressler}
{Dressler} A.,  1980, \mn@doi [\apj] {10.1086/157753}, \href {https://ui.adsabs.harvard.edu/abs/1980ApJ...236..351D} {236, 351}

\bibitem[\protect\citeauthoryear{{Dressler} et~al.,}{{Dressler} et~al.}{1997}]{1997ApJ...490..577D}
{Dressler} A.,  et~al., 1997, \mn@doi [\apj] {10.1086/304890}, \href {https://ui.adsabs.harvard.edu/abs/1997ApJ...490..577D} {490, 577}

\bibitem[\protect\citeauthoryear{{El-Badry}, {Wetzel}, {Geha}, {Hopkins}, {Kere{\v{s}}}, {Chan}  \& {Faucher-Gigu{\`e}re}}{{El-Badry} et~al.}{2016}]{2016ApJ...820..131E}
{El-Badry} K.,  {Wetzel} A.,  {Geha} M.,  {Hopkins} P.~F.,  {Kere{\v{s}}} D.,  {Chan} T.~K.,   {Faucher-Gigu{\`e}re} C.-A.,  2016, \mn@doi [\apj] {10.3847/0004-637X/820/2/131}, \href {https://ui.adsabs.harvard.edu/abs/2016ApJ...820..131E} {820, 131}

\bibitem[\protect\citeauthoryear{Faber}{Faber}{2011}]{faber2011cosmic}
Faber S.,  2011, MAST, doi, 10, T94S3X

\bibitem[\protect\citeauthoryear{Ferrari, de Carvalho  \& Trevisan}{Ferrari et~al.}{2015}]{ferrari2015morfometryka}
Ferrari F.,  de Carvalho R.~R.,   Trevisan M.,  2015, The Astrophysical Journal, 814, 55

\bibitem[\protect\citeauthoryear{{Fioc} \& {Rocca-Volmerange}}{{Fioc} \& {Rocca-Volmerange}}{1997}]{1997A&A...326..950F}
{Fioc} M.,  {Rocca-Volmerange} B.,  1997, \mn@doi [\aap] {10.48550/arXiv.astro-ph/9707017}, \href {https://ui.adsabs.harvard.edu/abs/1997A&A...326..950F} {326, 950}

\bibitem[\protect\citeauthoryear{Grogin et~al.,}{Grogin et~al.}{2011}]{grogin2011candels}
Grogin N.~A.,  et~al., 2011, The Astrophysical Journal Supplement Series, 197, 35

\bibitem[\protect\citeauthoryear{Hausen \& Robertson}{Hausen \& Robertson}{2020}]{hausen2020morpheus}
Hausen R.,  Robertson B.~E.,  2020, The Astrophysical Journal Supplement Series, 248, 20

\bibitem[\protect\citeauthoryear{{Hopkins}, {Hernquist}, {Cox}, {Di Matteo}, {Robertson}  \& {Springel}}{{Hopkins} et~al.}{2006}]{2006ApJS..163....1H}
{Hopkins} P.~F.,  {Hernquist} L.,  {Cox} T.~J.,  {Di Matteo} T.,  {Robertson} B.,   {Springel} V.,  2006, \mn@doi [\apjs] {10.1086/499298}, \href {https://ui.adsabs.harvard.edu/abs/2006ApJS..163....1H} {163, 1}

\bibitem[\protect\citeauthoryear{{Jackson}, {Martin}, {Kaviraj}, {Laigle}, {Devriendt}, {Dubois}  \& {Pichon}}{{Jackson} et~al.}{2019}]{2019MNRAS.489.4679J}
{Jackson} R.~A.,  {Martin} G.,  {Kaviraj} S.,  {Laigle} C.,  {Devriendt} J.~E.~G.,  {Dubois} Y.,   {Pichon} C.,  2019, \mn@doi [\mnras] {10.1093/mnras/stz2440}, \href {https://ui.adsabs.harvard.edu/abs/2019MNRAS.489.4679J} {489, 4679}

\bibitem[\protect\citeauthoryear{Jensen}{Jensen}{1996}]{jensen1996introductory}
Jensen J.~R.,  1996, Introductory digital image processing: a remote sensing perspective..
Taylor \& Francis

\bibitem[\protect\citeauthoryear{Kartaltepe et~al.,}{Kartaltepe et~al.}{2015}]{kartaltepe2015candels}
Kartaltepe J.~S.,  et~al., 2015, The Astrophysical Journal Supplement Series, 221, 11

\bibitem[\protect\citeauthoryear{Khalifa, Taha, Hassanien  \& Selim}{Khalifa et~al.}{2017}]{khalifa2017deep}
Khalifa N. E.~M.,  Taha M. H.~N.,  Hassanien A.~E.,   Selim I.,  2017, arXiv preprint arXiv:1709.02245

\bibitem[\protect\citeauthoryear{Khan, Huerta, Wang, Gruendl, Jennings  \& Zheng}{Khan et~al.}{2019}]{khan2019deep}
Khan A.,  Huerta E.,  Wang S.,  Gruendl R.,  Jennings E.,   Zheng H.,  2019, Physics Letters B, 795, 248

\bibitem[\protect\citeauthoryear{{Khochfar} \& {Burkert}}{{Khochfar} \& {Burkert}}{2003}]{2003ApJ...597L.117K}
{Khochfar} S.,  {Burkert} A.,  2003, \mn@doi [\apjl] {10.1086/379845}, \href {https://ui.adsabs.harvard.edu/abs/2003ApJ...597L.117K} {597, L117}

\bibitem[\protect\citeauthoryear{Koekemoer et~al.,}{Koekemoer et~al.}{2011}]{koekemoer2011candels}
Koekemoer A.~M.,  et~al., 2011, The Astrophysical Journal Supplement Series, 197, 36

\bibitem[\protect\citeauthoryear{{Kolesnikov}, {Sampaio}, {de Carvalho}, {Conselice}, {Rembold}, {Mendes}  \& {Rosa}}{{Kolesnikov} et~al.}{2024}]{kolesnikov2023unveiling}
{Kolesnikov} I.,  {Sampaio} V.~M.,  {de Carvalho} R.~R.,  {Conselice} C.,  {Rembold} S.~B.,  {Mendes} C.~L.,   {Rosa} R.~R.,  2024, \mn@doi [\mnras] {10.1093/mnras/stad3934}, \href {https://ui.adsabs.harvard.edu/abs/2024MNRAS.528...82K} {528, 82}

\bibitem[\protect\citeauthoryear{Lee, Park, Hwang  \& Kwon}{Lee et~al.}{2024}]{lee2024morphology}
Lee J.~H.,  Park C.,  Hwang H.~S.,   Kwon M.,  2024, The Astrophysical Journal, 966, 113

\bibitem[\protect\citeauthoryear{Lintott et~al.,}{Lintott et~al.}{2011}]{lintott2011galaxy}
Lintott C.,  et~al., 2011, Monthly Notices of the Royal Astronomical Society, 410, 166

\bibitem[\protect\citeauthoryear{Lloyd}{Lloyd}{1982}]{lloyd1982least}
Lloyd S.,  1982, IEEE transactions on information theory, 28, 129

\bibitem[\protect\citeauthoryear{Lotz, Primack  \& Madau}{Lotz et~al.}{2004}]{lotz2004new}
Lotz J.~M.,  Primack J.,   Madau P.,  2004, The Astronomical Journal, 128, 163

\bibitem[\protect\citeauthoryear{{Madau} \& {Dickinson}}{{Madau} \& {Dickinson}}{2014}]{2014ARA&A..52..415M}
{Madau} P.,  {Dickinson} M.,  2014, \mn@doi [\araa] {10.1146/annurev-astro-081811-125615}, \href {https://ui.adsabs.harvard.edu/abs/2014ARA&A..52..415M} {52, 415}

\bibitem[\protect\citeauthoryear{{Maraston}}{{Maraston}}{2005}]{2005MNRAS.362..799M}
{Maraston} C.,  2005, \mn@doi [\mnras] {10.1111/j.1365-2966.2005.09270.x}, \href {https://ui.adsabs.harvard.edu/abs/2005MNRAS.362..799M} {362, 799}

\bibitem[\protect\citeauthoryear{{Moody}, {Romanowsky}, {Cox}, {Novak}  \& {Primack}}{{Moody} et~al.}{2014}]{2014MNRAS.444.1475M}
{Moody} C.~E.,  {Romanowsky} A.~J.,  {Cox} T.~J.,  {Novak} G.~S.,   {Primack} J.~R.,  2014, \mn@doi [\mnras] {10.1093/mnras/stu1444}, \href {https://ui.adsabs.harvard.edu/abs/2014MNRAS.444.1475M} {444, 1475}

\bibitem[\protect\citeauthoryear{Mortlock et~al.,}{Mortlock et~al.}{2013}]{mortlock2013redshift}
Mortlock A.,  et~al., 2013, Monthly Notices of the Royal Astronomical Society, 433, 1185

\bibitem[\protect\citeauthoryear{Nayyeri et~al.,}{Nayyeri et~al.}{2017}]{nayyeri2017candels}
Nayyeri H.,  et~al., 2017, The Astrophysical Journal Supplement Series, 228, 7

\bibitem[\protect\citeauthoryear{{Peebles}}{{Peebles}}{1969}]{1969ApJ...155..393P}
{Peebles} P.~J.~E.,  1969, \mn@doi [\apj] {10.1086/149876}, \href {https://ui.adsabs.harvard.edu/abs/1969ApJ...155..393P} {155, 393}

\bibitem[\protect\citeauthoryear{{Planck Collaboration} et~al.,}{{Planck Collaboration} et~al.}{2020}]{2020A&A...641A...6P}
{Planck Collaboration} et~al., 2020, \mn@doi [\aap] {10.1051/0004-6361/201833910}, \href {https://ui.adsabs.harvard.edu/abs/2020A&A...641A...6P} {641, A6}

\bibitem[\protect\citeauthoryear{{Poggianti}}{{Poggianti}}{1997}]{1997A&AS..122..399P}
{Poggianti} B.~M.,  1997, \mn@doi [\aaps] {10.1051/aas:1997142}, \href {https://ui.adsabs.harvard.edu/abs/1997A&AS..122..399P} {122, 399}

\bibitem[\protect\citeauthoryear{Popp et~al.,}{Popp et~al.}{2024}]{popp2024transfer}
Popp J.~J.,  et~al., 2024, RAS Techniques and Instruments, 3, 174

\bibitem[\protect\citeauthoryear{Primack et~al.,}{Primack et~al.}{2018}]{primack2018deep}
Primack J.,  et~al., 2018, The Astrophysical Journal, 858, 114

\bibitem[\protect\citeauthoryear{Robertson et~al.,}{Robertson et~al.}{2023}]{robertson2023morpheus}
Robertson B.~E.,  et~al., 2023, The Astrophysical Journal Letters, 942, L42

\bibitem[\protect\citeauthoryear{{Rodriguez-Gomez} et~al.,}{{Rodriguez-Gomez} et~al.}{2015}]{2015MNRAS.449...49R}
{Rodriguez-Gomez} V.,  et~al., 2015, \mn@doi [\mnras] {10.1093/mnras/stv264}, \href {https://ui.adsabs.harvard.edu/abs/2015MNRAS.449...49R} {449, 49}

\bibitem[\protect\citeauthoryear{{Rodriguez-Gomez} et~al.,}{{Rodriguez-Gomez} et~al.}{2017}]{2017MNRAS.467.3083R}
{Rodriguez-Gomez} V.,  et~al., 2017, \mn@doi [\mnras] {10.1093/mnras/stx305}, \href {https://ui.adsabs.harvard.edu/abs/2017MNRAS.467.3083R} {467, 3083}

\bibitem[\protect\citeauthoryear{Rosa et~al.,}{Rosa et~al.}{2018}]{rosa2018gradient}
Rosa R.,  et~al., 2018, Monthly Notices of the Royal Astronomical Society: Letters, 477, L101

\bibitem[\protect\citeauthoryear{{Sampaio}, {de Carvalho}, {Ferreras}, {Arag{\'o}n-Salamanca}  \& {Parker}}{{Sampaio} et~al.}{2022}]{2022MNRAS.509..567S}
{Sampaio} V.~M.,  {de Carvalho} R.~R.,  {Ferreras} I.,  {Arag{\'o}n-Salamanca} A.,   {Parker} L.~C.,  2022, \mn@doi [\mnras] {10.1093/mnras/stab3018}, \href {https://ui.adsabs.harvard.edu/abs/2022MNRAS.509..567S} {509, 567}

\bibitem[\protect\citeauthoryear{{Sampaio}, {Arag{\'o}n-Salamanca}, {Merrifield}, {de Carvalho}, {Zhou}  \& {Ferreras}}{{Sampaio} et~al.}{2023}]{2023MNRAS.524.5327S}
{Sampaio} V.~M.,  {Arag{\'o}n-Salamanca} A.,  {Merrifield} M.~R.,  {de Carvalho} R.~R.,  {Zhou} S.,   {Ferreras} I.,  2023, \mn@doi [\mnras] {10.1093/mnras/stad2211}, \href {https://ui.adsabs.harvard.edu/abs/2023MNRAS.524.5327S} {524, 5327}

\bibitem[\protect\citeauthoryear{{Sampaio}, {de Carvalho}, {Arag{\'o}n-Salamanca}, {Merrifield}, {Ferreras}  \& {Cornwell}}{{Sampaio} et~al.}{2024}]{2024MNRAS.532..982S}
{Sampaio} V.~M.,  {de Carvalho} R.~R.,  {Arag{\'o}n-Salamanca} A.,  {Merrifield} M.~R.,  {Ferreras} I.,   {Cornwell} D.~J.,  2024, \mn@doi [\mnras] {10.1093/mnras/stae1533}, \href {https://ui.adsabs.harvard.edu/abs/2024MNRAS.532..982S} {532, 982}

\bibitem[\protect\citeauthoryear{{Sandage}}{{Sandage}}{1961}]{1961ApJ...134..916S}
{Sandage} A.,  1961, \mn@doi [\apj] {10.1086/147218}, \href {https://ui.adsabs.harvard.edu/abs/1961ApJ...134..916S} {134, 916}

\bibitem[\protect\citeauthoryear{Santini et~al.,}{Santini et~al.}{2015}]{santini2015stellar}
Santini P.,  et~al., 2015, The Astrophysical Journal, 801, 97

\bibitem[\protect\citeauthoryear{{Schawinski} et~al.,}{{Schawinski} et~al.}{2014}]{2014MNRAS.440..889S}
{Schawinski} K.,  et~al., 2014, \mn@doi [\mnras] {10.1093/mnras/stu327}, \href {https://ui.adsabs.harvard.edu/abs/2014MNRAS.440..889S} {440, 889}

\bibitem[\protect\citeauthoryear{{Simpson}, {Bryan}, {Johnston}, {Smith}, {Mac Low}, {Sharma}  \& {Tumlinson}}{{Simpson} et~al.}{2013}]{2013MNRAS.432.1989S}
{Simpson} C.~M.,  {Bryan} G.~L.,  {Johnston} K.~V.,  {Smith} B.~D.,  {Mac Low} M.-M.,  {Sharma} S.,   {Tumlinson} J.,  2013, \mn@doi [\mnras] {10.1093/mnras/stt474}, \href {https://ui.adsabs.harvard.edu/abs/2013MNRAS.432.1989S} {432, 1989}

\bibitem[\protect\citeauthoryear{Simpson, Page  \& De~Roure}{Simpson et~al.}{2014}]{simpson2014zooniverse}
Simpson R.,  Page K.~R.,   De~Roure D.,  2014, in Proceedings of the 23rd international conference on world wide web. pp 1049--1054

\bibitem[\protect\citeauthoryear{{Springel} \& {Hernquist}}{{Springel} \& {Hernquist}}{2005}]{2005ApJ...622L...9S}
{Springel} V.,  {Hernquist} L.,  2005, \mn@doi [\apjl] {10.1086/429486}, \href {https://ui.adsabs.harvard.edu/abs/2005ApJ...622L...9S} {622, L9}

\bibitem[\protect\citeauthoryear{Stefanon et~al.,}{Stefanon et~al.}{2017}]{stefanon2017candels}
Stefanon M.,  et~al., 2017, The Astrophysical Journal Supplement Series, 229, 32

\bibitem[\protect\citeauthoryear{{Strateva} et~al.,}{{Strateva} et~al.}{2001}]{Strateva1}
{Strateva} I.,  et~al., 2001, \mn@doi [\aj] {10.1086/323301}, \href {https://ui.adsabs.harvard.edu/abs/2001AJ....122.1861S} {122, 1861}

\bibitem[\protect\citeauthoryear{Talia, Cimatti, Mignoli, Pozzetti, Renzini, Kurk  \& Halliday}{Talia et~al.}{2014}]{talia2014listening}
Talia M.,  Cimatti A.,  Mignoli M.,  Pozzetti L.,  Renzini A.,  Kurk J.,   Halliday C.,  2014, Astronomy \& Astrophysics, 562, A113

\bibitem[\protect\citeauthoryear{Tasca et~al.,}{Tasca et~al.}{2009}]{tasca2009zcosmos}
Tasca L.~A.,  et~al., 2009, Astronomy \& Astrophysics, 503, 379

\bibitem[\protect\citeauthoryear{{Teklu}, {Remus}, {Dolag}, {Beck}, {Burkert}, {Schmidt}, {Schulze}  \& {Steinborn}}{{Teklu} et~al.}{2015}]{2015ApJ...812...29T}
{Teklu} A.~F.,  {Remus} R.-S.,  {Dolag} K.,  {Beck} A.~M.,  {Burkert} A.,  {Schmidt} A.~S.,  {Schulze} F.,   {Steinborn} L.~K.,  2015, \mn@doi [\apj] {10.1088/0004-637X/812/1/29}, \href {https://ui.adsabs.harvard.edu/abs/2015ApJ...812...29T} {812, 29}

\bibitem[\protect\citeauthoryear{Tohill, Ferreira, Conselice, Bamford  \& Ferrari}{Tohill et~al.}{2021}]{tohill2021quantifying}
Tohill C.,  Ferreira L.,  Conselice C.,  Bamford S.,   Ferrari F.,  2021, The Astrophysical Journal, 916, 4

\bibitem[\protect\citeauthoryear{{Toomre} \& {Toomre}}{{Toomre} \& {Toomre}}{1972}]{1972ApJ...178..623T}
{Toomre} A.,  {Toomre} J.,  1972, \mn@doi [\apj] {10.1086/151823}, \href {https://ui.adsabs.harvard.edu/abs/1972ApJ...178..623T} {178, 623}

\bibitem[\protect\citeauthoryear{{Trussler}, {Maiolino}, {Maraston}, {Peng}, {Thomas}, {Goddard}  \& {Lian}}{{Trussler} et~al.}{2020}]{2020MNRAS.491.5406T}
{Trussler} J.,  {Maiolino} R.,  {Maraston} C.,  {Peng} Y.,  {Thomas} D.,  {Goddard} D.,   {Lian} J.,  2020, \mn@doi [\mnras] {10.1093/mnras/stz3286}, \href {https://ui.adsabs.harvard.edu/abs/2020MNRAS.491.5406T} {491, 5406}

\bibitem[\protect\citeauthoryear{Velasco}{Velasco}{1979}]{velasco1979thresholding}
Velasco F.~R.,  1979, Thresholding using the ISODATA clustering algorithm.
University of Maryland, Computer Science Center

\bibitem[\protect\citeauthoryear{Villa-Vialaneix}{Villa-Vialaneix}{2017}]{villa2017stochastic}
Villa-Vialaneix N.,  2017, in 2017 12th International Workshop on Self-Organizing Maps and Learning Vector Quantization, Clustering and Data Visualization (WSOM). pp~1--7

\bibitem[\protect\citeauthoryear{Walmsley et~al.,}{Walmsley et~al.}{2022}]{walmsley2022galaxy}
Walmsley M.,  et~al., 2022, Monthly Notices of the Royal Astronomical Society, 509, 3966

\bibitem[\protect\citeauthoryear{Walmsley et~al.,}{Walmsley et~al.}{2023}]{walmsley2023galaxy}
Walmsley M.,  et~al., 2023, Monthly Notices of the Royal Astronomical Society, 526, 4768

\bibitem[\protect\citeauthoryear{Wei, Lu, Dai, Liang, Hao, Zhang  \& Zhang}{Wei et~al.}{2023}]{wei2023galaxy}
Wei S.,  Lu W.,  Dai W.,  Liang B.,  Hao L.,  Zhang Z.,   Zhang X.,  2023, The Astronomical Journal, 167, 29

\bibitem[\protect\citeauthoryear{{Wetzel}, {Tinker}  \& {Conroy}}{{Wetzel} et~al.}{2012}]{2012MNRAS.424..232W}
{Wetzel} A.~R.,  {Tinker} J.~L.,   {Conroy} C.,  2012, \mn@doi [\mnras] {10.1111/j.1365-2966.2012.21188.x}, \href {https://ui.adsabs.harvard.edu/abs/2012MNRAS.424..232W} {424, 232}

\bibitem[\protect\citeauthoryear{Whitney, Ferreira, Conselice  \& Duncan}{Whitney et~al.}{2021}]{whitney2021galaxy}
Whitney A.,  Ferreira L.,  Conselice C.,   Duncan K.,  2021, The Astrophysical Journal, 919, 139

\bibitem[\protect\citeauthoryear{{York} et~al.,}{{York} et~al.}{2000}]{2000AJ....120.1579Y}
{York} D.~G.,  et~al., 2000, \mn@doi [\aj] {10.1086/301513}, \href {https://ui.adsabs.harvard.edu/abs/2000AJ....120.1579Y} {120, 1579}

\bibitem[\protect\citeauthoryear{{Zeng}, {Wang}, {Gao}  \& {Yang}}{{Zeng} et~al.}{2024}]{2024MNRAS.532.2558Z}
{Zeng} G.,  {Wang} L.,  {Gao} L.,   {Yang} H.,  2024, \mn@doi [\mnras] {10.1093/mnras/stae1651}, \href {https://ui.adsabs.harvard.edu/abs/2024MNRAS.532.2558Z} {532, 2558}

\bibitem[\protect\citeauthoryear{Zhao, Alzubaidi, Zhang, Duan  \& Gu}{Zhao et~al.}{2023}]{zhao2023comparison}
Zhao Z.,  Alzubaidi L.,  Zhang J.,  Duan Y.,   Gu Y.,  2023, Expert Systems with Applications, p. 122807

\bibitem[\protect\citeauthoryear{van~der Wel et~al.,}{van~der Wel et~al.}{2014}]{van20143d}
van~der Wel A.,  et~al., 2014, The Astrophysical Journal, 788, 28

\makeatother
\end{thebibliography}


\section{Appendix A: Fractions of the different morphological types as presented in Figures 10 and 11}\label{sec:appendix1}

Appendix provide fractions provides earlier in tabular format for readers convenience.

\begin{table}
\caption{Table of fractions corresponding to Figure \ref{fig:frac_main}. For each redshift bin ($z$), we provide the fractions of disk (D), spheroidal (S), and irregular (I) galaxies, along with their respective uncertainties, shown in the $\rm Q_{\sigma_{D}}$, $\rm Q_{\sigma_{S}}$, and $\rm Q_{\sigma_{I}}$ columns.}
\begin{tabular}{|c|c|c|c|c|c|c|}
\hline
\textbf{$z$} & \textbf{D} & \textbf{$\rm Q_{\sigma_{D}}$} & \textbf{S} & \textbf{$\rm Q_{\sigma_{S}}$} & \textbf{I} & \textbf{$\rm Q_{\sigma_{I}}$} \\ \hline
0.3       & 0.53       & 0.09          & 0.33       & 0.08          & 0.14       & 0.02          \\ \hline
0.5       & 0.60       & 0.07          & 0.30       & 0.04          & 0.10       & 0.02          \\ \hline
0.7       & 0.58       & 0.05          & 0.34       & 0.05          & 0.08       & 0.01          \\ \hline
0.9       & 0.60       & 0.05          & 0.32       & 0.04          & 0.08       & 0.01          \\ \hline
1.1       & 0.54       & 0.03          & 0.39       & 0.03          & 0.06       & 0.01          \\ \hline
1.3       & 0.52       & 0.06          & 0.38       & 0.05          & 0.11       & 0.01          \\ \hline
1.5       & 0.59       & 0.04          & 0.31       & 0.03          & 0.10       & 0.01          \\ \hline
1.7       & 0.51       & 0.05          & 0.36       & 0.05          & 0.13       & 0.01          \\ \hline
1.9       & 0.65       & 0.03          & 0.25       & 0.03          & 0.10       & 0.01          \\ \hline
2.1       & 0.50       & 0.05          & 0.33       & 0.06          & 0.17       & 0.02          \\ \hline
2.3       & 0.60       & 0.05          & 0.24       & 0.04          & 0.16       & 0.02          \\ \hline
\end{tabular}

\label{table:main_frac}
\end{table}

\begin{table}
\caption{Table of fractions corresponding to the Figure \ref{fig:super_lee}.  For each redshift range ($z$) and corresponding Stellar Mass bin ($\rm \log M_{\rm stellar} \, ({\rm M}_{\odot})$) we provide the fractions of disk (D), spheroidal (S), and irregular (I) galaxies, along with their respective uncertainties, shown in the $\rm Q_{\sigma_{D}}$, $\rm Q_{\sigma_{S}}$, and $\rm Q_{\sigma_{I}}$ columns.}
\begin{tabular}{|c|c|c|c|c|c|c|c|}
\hline
\textbf{$z$}             & \textbf{$\rm \log M_{\rm stellar} \, ({\rm M}_{\odot})$} & \textbf{D} & \textbf{$\rm Q_{\sigma_{D}}$} & \textbf{S} & \textbf{$\rm Q_{\sigma_{S}}$} & \textbf{I} & \textbf{$\rm Q_{\sigma_{I}}$} \\ \hline
$0.6 < z < 1.2$ & 9.00          & 0.70       & 0.03        & 0.23       & 0.03        & 0.08       & 0.01         \\ \hline
$0.6 < z < 1.2$ & 9.75          & 0.62       & 0.04        & 0.29       & 0.03        & 0.09       & 0.01         \\ \hline
$0.6 < z < 1.2$ & 10.50         & 0.37       & 0.03        & 0.57       & 0.03        & 0.07       & 0.01         \\ \hline
$0.6 < z < 1.2$ & 11.25         & 0.15       & 0.02        & 0.79       & 0.03        & 0.06       & 0.01         \\ \hline
$0.6 < z < 1.2$ & 12.00         & 0.00       & 0.00        & 1.00       & 0.06        & 0.00       & 0.06         \\ \hline

$1.2 < z < 2.0$ & 9.00          & 0.56       & 0.03              & 0.32       & 0.03              & 0.12       & 0.01              \\ \hline
$1.2 < z < 2.0$ & 9.75          & 0.63       & 0.03              & 0.26       & 0.02              & 0.11       & 0.01              \\ \hline
$1.2 < z < 2.0$ & 10.50         & 0.52       & 0.03              & 0.38       & 0.03              & 0.09       & 0.01              \\ \hline
$1.2 < z < 2.0$ & 11.25         & 0.32       & 0.02              & 0.59       & 0.04              & 0.09       & 0.02              \\ \hline
$1.2 < z < 2.0$ & 12.00         & 0.22       & 0.01              & 0.70       & 0.04              & 0.09       & 0.06              \\ \hline

$2.0 < z < 2.4$ & 9.00          & 0.53       & 0.03              & 0.32       & 0.04              & 0.16       & 0.02              \\ \hline
$2.0 < z < 2.4$ & 9.75          & 0.57       & 0.03              & 0.27       & 0.03              & 0.16       & 0.02              \\ \hline
$2.0 < z < 2.4$ & 10.50         & 0.55       & 0.03              & 0.26       & 0.03              & 0.19       & 0.02              \\ \hline
$2.0 < z < 2.4$ & 11.25         & 0.44       & 0.04              & 0.41       & 0.06              & 0.15       & 0.04              \\ \hline
$2.0 < z < 2.4$ & 12.00         & 0.46       & 0.03              & 0.53       & 0.00              & 0.01       & 0.00              \\ \hline
\end{tabular}

\label{table:super_lee_fractions1}
\end{table}


\bsp	
\label{lastpage}
\end{document}